\documentclass[aps,prb,showpacs,superscriptaddress,reprint]{revtex4-1}

\usepackage{graphics}
\usepackage{graphicx}
\usepackage{epsfig}
\usepackage{amssymb}
\usepackage{amsthm}
\usepackage{xcolor}
\usepackage{lipsum}

\definecolor{GreenMarker}{rgb}{0.0, 0.75, 0.0}
\definecolor{BlueMarker}{rgb}{0.0, 0.0, 0.75}

\usepackage{amsmath}
\DeclareMathOperator{\arccosh}{acosh}

\begin{document}

\title{Magnetic flux penetration in nanoscale wedge-shaped superconducting thin films}

\author{L. B. L. G. Pinheiro}
\affiliation{Departamento de F\'{\i}sica, Universidade Federal de S\~{a}o Carlos, 13565-905 S\~{a}o Carlos, SP, Brazil}
\affiliation{Instituto Federal Federal de S\~{a}o Paulo, 13565-905 S\~{a}o Carlos, SP, Brazil} 

\author{L. Jiang}
\affiliation{School of Aeronautics, Northwestern Polytechnical University, Xi'an 710072, China}
\affiliation{D\'epartement de Physique, Universit\'e de Li\`ege, B-4000 Sart Tilman, Belgium}

\author{E. A. Abbey}
\affiliation{Departamento de F\'{\i}sica, Universidade Federal de S\~{a}o Carlos, 13565-905 S\~{a}o Carlos, SP, Brazil}

\author{Davi A. D. Chaves}
\affiliation{Departamento de F\'{\i}sica, Universidade Federal de S\~{a}o Carlos, 13565-905 S\~{a}o Carlos, SP, Brazil}


\author{A. J. Chiquito}
\affiliation{Departamento de F\'{\i}sica, Universidade Federal de S\~{a}o Carlos, 13565-905 S\~{a}o Carlos, SP, Brazil}

\author{T. H. Johansen}
\affiliation{Department of Physics, University of Oslo, P.O. Box 1048 Blindern, 0316 Oslo, Norway}

\author{J. Van de Vondel}
\affiliation{Quantum Solid State Physics, Department of Physics and Astronomy, KU Leuven, Celestijnenlaan 200D, B-3001 Leuven, Belgium}

\author{C. Xue}
\affiliation{School of Mechanics, Civil Engineering and Architecture, Northwestern Polytechnical University, Xi'an 710072, China}

\author{Y.-H. Zhou}
\affiliation{School of Aeronautics, Northwestern Polytechnical University, Xi'an 710072, China}
\affiliation{Department of Mechanics and Engineering Sciences, Lanzhou University, Key Laboratory of Mechanics on Disaster and Environment in Western China, Ministry of Education of China, Lanzhou 730000, China}

\author{A. V. Silhanek}
\affiliation{D\'epartement de Physique, Universit\'e de Li\`ege, B-4000 Sart Tilman, Belgium}

\author{W. A. Ortiz}
\affiliation{Departamento de F\'{\i}sica, Universidade Federal de S\~{a}o Carlos, 13565-905 S\~{a}o Carlos, SP, Brazil}

\author{M. Motta}
\email[Corresponding author: ]{m.motta@df.ufscar.br}
\affiliation{Departamento de F\'{\i}sica, Universidade Federal de S\~{a}o Carlos, 13565-905 S\~{a}o Carlos, SP, Brazil}

\date{\today} 

\begin{abstract}

Thickness uniformity is regarded as an important parameter in designing thin film devices. However, some applications based on films with non-uniform thickness have recently emerged, such as gas sensors and optimized materials based on the gradual change of film composition. This work deals with superconducting Pb thin films with a thickness gradient prepared with the aid of a diffuse stencil mask. Atomic Force Microscopy and Energy-Dispersive X-ray Spectroscopy show variations ranging from 90~nm to 154~nm. Quantitative magneto-optical images reveal interesting features during both the abrupt and the smooth penetration regimes of magnetic flux, as well as the thickness-dependent critical current density ($J_c$). In addition, we observe a gradual superconducting transition as the upper critical field is progressively reached for certain thicknesses. Furthermore, the hysteresis observed for triggering flux avalanches when increasing and decreasing magnetic fields is also accounted for by the $J_c$ profile evolution along the thickness gradient. Numerical simulations based on the Thermomagnetic Model are in fair agreement with the experimental data.
These findings demonstrate that wedge-shaped films are a viable approach to investigate, in a continuous fashion, thickness-dependent properties of a superconducting materials.

\end{abstract}


\maketitle


\section{Introduction}

Several promising technologies, such as superconducting qubits~\cite{Wendin2017,kjaergaard2020superconducting} or single photon detectors~\cite{Hadfield2009,Natarajan2012}, are based on thin films. Non-uniform thickness distributions are described as a common point of concern for different evaporation techniques~\cite{SamiFranssila2010}. In fact, great effort has been exerted to optimize preparation conditions to guarantee uniformity and reproducibility in films used for a large diversity of applications~\cite{Luna2007unif,Yamamura2008unif,Choi2010unif,Wang2018unif,Kim2018unif,Knehr2021unif}. However, it is possible to harness differences in material properties arising from thickness gradients to develop devices with complex responses. For example, the ideal composition of Ti-Ni-based shape memory alloys with near-zero thermal hysteresis was identified and latter produced as a bulky material following an approach based on thin-film combinatorial deposition of nanoscale wedge-like multilayers with varying material composition and high-throughput characterization methods~\cite{Zarnetta2010,Ludwig2019}. 
Recently, this strategy was also applied to control the Ce doping concentration in the superconductor La$_{2-x}$Ce$_x$CuO$_4$ to investigate the anomalous non-Fermi liquid behavior and the mechanism for the superconductivity in those systems~\cite{Yuan2022}. Another example is related to novel gas sensors using thickness gradient films to detect low concentrations of hydrogen and ethanol gases~\cite{Palmisano2010,Shimizu2021}.
Furthermore, wedge-shaped superconducting-ferromagnetic heterostructures have successfully been used to prepare tunnel junction arrays presenting multiple 0-$\pi$ transitions~\cite{Born2006} and spin-valve cores~\cite{Antropov2013}.


For single superconducting layers, specimens with graded thickness or non-uniform critical current density, $J_c$, have been investigated on a theoretical and numerical basis~\cite{Du1995,Chapman1996,Sardella2009,LuZhou2016}. Exceptionally, Sabatino \textit{et al.}~\cite{Sabatino2012} unveiled a directed vortex motion on a uniform film of micrometric dimensions with asymmetric thickness profile edges and confirmed their findings by time-dependent Ginzburg-Landau (TDGL) simulations. Later on, Gladilin \textit{et al.}~\cite{Gladilin2015} employed the TDGL formalism to numerically investigate the dynamics of penetrated magnetic flux in a wedge-shaped film made of a type-I superconductor and steep enough to allow the thinnest part to behave as a type-II superconductor. In general, for non-uniform thick samples, there are a number of numerical and experimental investigations pointing to an important thickness dependent $J_c(d)$ ~\cite{Hengstberger2010,Mogro-Campero1990,Foltyn2003,Onori1985,Chaudhari1965,Ilin2010,Talantsev2015,Brisbois2017}.

For type-II superconductors, critical state models~\cite{bean1962prl,Kim1963,Fietz1964} are a powerful tool to unveil superconducting properties. They state that current flows in the superconductor at its critical value wherever there are vortices penetrated. For the simpler Bean model~\cite{bean1962prl}, $J_c$ is independent of the local flux density, $B_z$, however, a flux-dependent $J_c(B_z)$ may be essential to describe some subtle features~\cite{burger2019numerical,jiang2020selective,motta2021metamorphosis,davi2021}. In general, knowledge of the flux distribution provides information about the current distribution throughout the specimen~\cite{Brandt1997,Brandt1994,Clem1994,Brandt1994a,zeldov1994profile}. The critical state models do not account for all possible scenarios of magnetic flux penetration. This is particularly the case for thin films, for which stochastic abrupt dendritic flux penetration events, known as flux avalanches, take place at low temperatures~\cite{altshuler_colloquium:_2004,Denisov2006,colauto2020controlling}. The origin of such avalanches are thermomagnetic instabilities~\cite{Denisov2006}, occurring when the material cannot assimilate the heat generated by vortex motion, thus resulting in a positive-gain feedback loop which initiates a macroscopic magnetic flux burst characterized by a dendritic pattern.


Despite the unpredictable nature of these catastrophic events, much is known about the general influence of different environmental and sample parameters on their behavior~\cite{Yurchenko2009,vestgaarden2018nucleation}. For instance, an increase in temperature leads to fewer but more branched and larger avalanches~\cite{Vestgarden2011,blanco2019statistics}. They are also triggered for lower fields in films decorated with microscopic arrays of defects~\cite{Menghini2005,colauto2020controlling}. Besides that, their morphology is profoundly influenced by the geometry of the defect and its lattice symmetry~\cite{Motta2014}. Film thickness also influences the flux avalanches since thinner samples require a lower applied magnetic field to trigger the first avalanche~\cite{Abalosheva2010,Vestgarden2013arxiv,Abaloszewa2022}. These abrupt flux penetrations may be detrimental to applications because an avalanche can locally destroy the superconducting state~\cite{vestgaarden2018nucleation}. As such, their understanding is crucial for developing large-area thin superconducting devices. In particular, a comprehensive experimental study about the effects of thickness variation throughout large-areas is still lacking.

In this work, we report experimental and numerical investigations on the flux penetration in a type-II superconducting Pb wedge-shaped film of millimetric lateral dimensions. Atomic Force Microscopy (AFM), Energy-Dispersive X-ray Spectroscopy (EDS) and Scanning Electron Microscopy (SEM) were used to characterize the wedge profile and the structural properties of the sample. The Magneto-optical Imaging (MOI) technique and numerical simulations based on the Thermomagnetic (TM) model shed light on the magnetic flux penetration and the superconducting current distribution as the thickness changes gradually throughout the specimen. Both MOI and TM model reveal the impact of thickness on the avalanche morphology and their dependence on the magnetic history.



\section{Materials and Methods}

\subsection{Sample details}

A $2\times 7$~mm$^{2}$ Pb film was deposited onto a Si (100) substrate by conventional thermal evaporation with a thickness gradient along its longest dimension. A diffuse-shadow mask was used to create the wedge, spanning the thickness from 90 nm to 154 nm. A 20 nm-thick uniform protective Ge layer was deposited on top of the film. Microstructural characterization was done by AFM and SEM/EDS. The AFM images were captured in the peak-force tapping mode using a Digital Instruments Nanoscope V. The SEM/EDS measurements were carried out in a Philips XL-30 FEG Scanning Electron Microscope together with a XFLash 6/60 Bruker X-ray detector. Details of the sample fabrication and SEM/EDS analyses are presented in Appendices~\ref{sec:SampleDeposition} and \ref{sec:EDS}, respectively.



\subsection{Quantitative MOI}

The MOI technique is employed to visualize the spatial distribution of penetrated flux in the nanoscale wedge-shaped superconducting film. This technique relies on the Faraday effect in an indicator film, where polarized light will have its polarization rotated proportionally to the local magnetic field in the indicator, placed on top of the superconducting specimen. The indicator used in the present work is a Bi-substituted yttrium iron garnet film (Bi:YIG) with in-plane magnetization, covered with a 100 nm-thick Al mirror~\cite{Helseth2001,Helseth2002}. The resulting magneto-optical (MO) images are captured by a CCD (Charge-Couple Device) camera and show a qualitative magnetic flux distribution throughout the superconducting film, where the local brightness is related to the magnitude of the perpendicular flux density. A quantitative $B_z(x,y)$ picture is obtained using a pixel-by-pixel calibration algorithm implemented on MATLAB~\cite{shaw2018quantitative}. We also correct for sample drift with a precision of $\pm2$ pixels (or $\pm 8.3$~$\mu$m) in the position of any given image during the runs by employing the plugin StackReg~\cite{stackreg} together with ImageJ software~\cite{schneider2012nih}. The two-dimensional current distribution can be obtained from $B_z(x,y)$ by making a numerical inversion of the Biot-Savart law. In our case, we used the algorithm provided by Meltzer \textit{et al.}~\cite{meltzer2017reconstruction}.



The critical temperature of a sister sample from the same batch was determined in a Quantum Design MPMS-5S magnetometer using AC susceptibility measurements is $T_c$ = (7.20 $\pm$ 0.05)~K, whereas $T_c$ determined by the lack of image contrast using the MOI is $T_c^{MOI}$~=~(5.1 $\pm$ 0.1)~K. This difference is a consequence of a non-optimum thermal contact between the sample and the cold-finger. In order to obtain a meaningful comparison between MOI data and numerical simulations, the former is expressed in units of reduced temperature $T/T_c^{MOI}$ and the latter in $t=T/T_c$. 

\subsection{Numerical Simulations}

The magnetic flux penetration in the wedge-shaped superconducting Pb films is further numerically investigated by the TM model~\cite{Denisov2006,Denisov2006a,Vestgarden2011}, providing deeper insight into the observed thickness-dependent characteristic of the flux instabilities. Here, we consider a superconducting film with length of $2L=7$ mm and width of $2W=2$ mm, the same as the deposited film, and space-varying thickness $d(x,y)$. The sample is zero-field cooled (ZFC) to a base temperature $T_{0}$. The nonlinear material characteristics of the superconductor are described by the general $E$-$j$ constitutive law:
\begin{equation} \label{E1}
\textbf{E}=\left\{
\begin{array}{ll}
\rho_{n}(j/dJ_{c})^{n-1}\textbf{j}/d &\text{if}~j<dJ_{c}~\text{and}~T<T_{c},\\
\rho_{n}\textbf{j}/d &\text{otherwise}
\end{array}
\right.
\end{equation}
where $\rho_{n}$ is the normal state resistivity, $\textbf{j}$ is the sheet current which is defined as $\mathbf{j}(x,y)=\int_{-d/2}^{d/2}\mathbf{J}(x,y,z)dz$, and $n$ is the flux creep exponent. The temperature dependencies are $J_{c}=J_{c0}(d)(1-T/T_c)$ and $n=n_{0}T_{c}/T-50$~\cite{Motta2014}. The $J_c$ dependency on thickness across the sample has been taken from the experimental results as will be explained in the next section.

The electrodynamics of the superconducting thin film exposed to a transverse magnetic field follows Maxwell's equations:
\begin{equation} \label{E2}
\dot{\textbf{B}}=-\nabla \times \textbf{E},~{\nabla} \times \textbf{H}=\textbf{j}\delta(z) \; \text{and} \; {\nabla} \cdot \textbf{B}=0,
\end{equation}
with $\textbf{B}=\mu_{0}\textbf{H}$, and ${\nabla} \cdot \textbf{j}=0$. Thermomagnetic instabilities resulting from coupling the nonlinear electromagnetic characteristics of superconductors and the Joule heating created by magnetic flux motion. Thus the electrodynamics is supplemented by the heat diffusion equation
\begin{equation} \label{E3}
dc\dot{T}=d\kappa\nabla^2 T-h(T-T_{0})+\textbf{j}\cdot\textbf{E},
\end{equation}
where, $\kappa$, $c$, and $h$ are the thermal conductivity, the specific heat, and the coefficient for heat removal to the substrate, which are all assumed to be proportional to $T^{3}$. The terms on the right side of Eq.~\ref{E3} are related to the heat conduction within the film, heat flow to the substrate, and positive feedback due to Joule heating, respectively.
A flux avalanche will occur if the superconductor cannot evacuate heat fast enough, triggering a positive-gain feedback loop depending on the relative significance between the magnetic flux diffusion ($D_m$) and the thermal diffusion ($D_t$) coefficients~\cite{mints_critical_1981}. The key parameter controlling the occurrence of avalanches is given by $\tau = D_t/D_m = \mu_0\kappa_0\sigma/c$.

We solve Eq.~\ref{E3} together with Eq.~\ref{E1} and Eq.~\ref{E2}, using a real space/Fourier space hybrid method proposed by Vestg\aa rden \textit{et al.}~\cite{Vestgarden2011} with boundary conditions with $j=0$ outside the superconducting film. The parameters used in the simulation are related to Pb films~\cite{Portela2015}: $T_c=7.2$ K, $\rho_n=5.7\times10^{-9}$ $\Omega$m, $\kappa(T=T_c)=20$ W/Km, $c(T=T_c)=3\times10^4$ J/Km$^3$, and $h(T=T_c)=1\times10^4$ W/Km$^2$. We choose $n_{0}=90$ and limit the creep exponent to a maximum value of $n_{max}=100$.

\section{Results and Discussion}

\begin{figure}[t]
\centering
\includegraphics[width=1\linewidth]{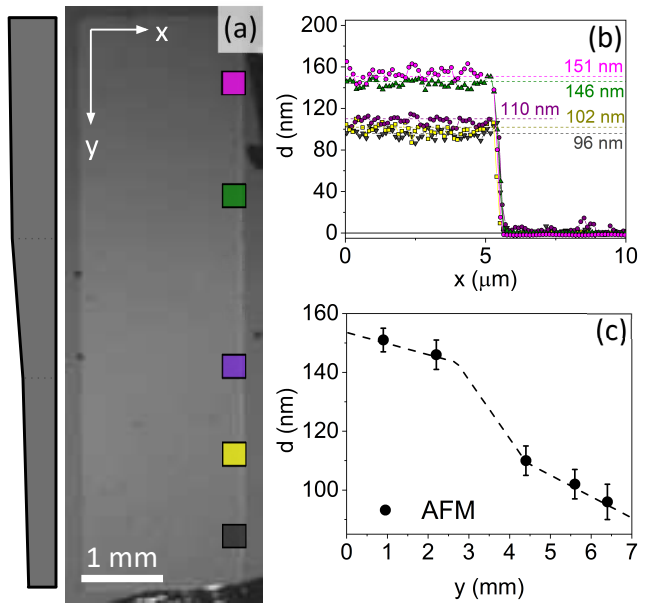}
\caption{Sample characterization. (a) Optical image of the wedge-shaped film, which shows five regions along the y axis where the thickness was determined. (b) Representative thickness profiles for the different indicated sample regions taken from AFM images. Dashed lines correspond to the averaged thickness over different profiles. (c) Thickness versus y position obtained from the AFM data. The dashed black line is the thickness used in the numerical simulations. The gray-black vertical bar at the left side of panel (a) is a pictorial representation of the thickness variation along the wedge.}
\label{fig_sample}
\end{figure}

An important first step consists in characterizing the structure of the wedge-shaped films from milimetric down to nanoscopic scales.  Fig.~\ref{fig_sample}(a) shows an optical image of the rectangular film with reasonably well-defined edges and the five regions where the thickness variation was investigated, indicated by colored square boxes. Representative thickness profiles obtained from AFM images are presented in panel (b). The thickness values for each region, represented by the colored dashed lines, were evaluated by averaging six different profiles and present standard deviations between 4~nm and 6~nm. To evaluate these Pb thicknesses, we subtract 20~nm referent to the Ge layer. Fig.~\ref{fig_sample}(c) shows the film thickness versus vertical position obtained from the AFM data. The dashed lines are an approximation of the real thickness profile and were used in the numerical simulations. The microstructure and thickness evolution throughout the film are corroborated by SEM images and EDS spectra, as shown in Appendix~\ref{sec:EDS}.

\begin{figure}[t]
\centering
\includegraphics[width=1\linewidth]{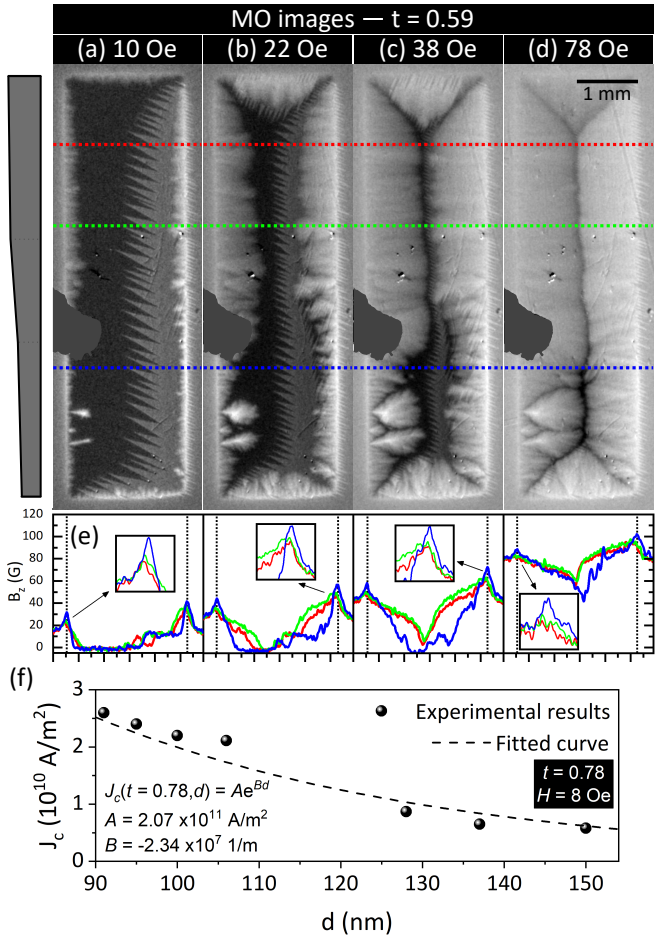}
\caption{MO images of the smooth flux penetration for wedge-shaped rectangular film captured at t = 0.59, after ZFC and for increasing H: (a) 10~Oe, (b) 22~Oe, (c) 38~Oe, and (d) 78~Oe. To avoid overexposure, each image was adjusted to have the best contrast. (e) $B_z$ profiles taken along the colored dashed lines in the MO images. (f) Thickness-dependent critical current density at $t$~=~0.78 and $H$~=~8~Oe, obtained by the flux front position along vertical left edge using the approximation given by Eq.~\eqref{Jc-Beanmodel}. The dashed line is an exponential fit of the experimental data, with parameters presented in the figure.}
\label{fig_MOI_smooth}
\end{figure}

The flux distribution along the gradient film was investigated by MOI, as shown in Fig.~\ref{fig_MOI_smooth}(a)-(d) at constant temperature $t$ = 0.59 and for increasing applied fields after a ZFC procedure. The gray smudge at the left side represents a region where vacuum grease unexpectedly jumped from the cold finger assembly during the experiment. Its gray level was kept constant for all MO images throughout the text. Fortunately, the smudge lies above the MO indicator and does not affect the response of the sample. The flux penetration observed at this temperature is as described by the critical state models~\cite{bean1962prl,bean1964rmp,Kim1963}.
In the top thicker half of the sample, the flux front penetrates deeper and becomes more apparent above $H$ = 22~Oe. Consequently, its shielding capability is inferior to that of the bottom half (thinner region). 
In addition, the flux fronts on the left and right borders close to the top and bottom edges are affected as a consequence of the supercurrents which have to adapt to the rectangular geometry of the film, resulting in black diagonal lines, also known as discontinuity lines or d-lines~\cite{Schuster1996}. The vertical d-line emerges gradually when the field reaches the center of the film, even though the flux front depth is not homogeneous and the full penetration field is not unique. The panels (a)-(d) of Fig.~\ref{fig_MOI_smooth} were selected to highlight the difference among the thicker and the thinner regions of the specimen.
The fact that we do not observe additional symmetrical d-lines anywhere in the film~\cite{Motta2014,valerio2021superconducting} is a strong indication that there exists a smooth thickness gradient, as represented in Fig.~\ref{fig_sample}(c). It is important mentioning that two flourishing patterns on the bottom left edge are due to defects which favor the flux penetration around them~\cite{Brisbois2016} and produce a minor distortion of the central d-line and a slight asymmetry of the side borders. This type of non-uniform flux penetration due to a thickness variation has been also observed in a V$_3$Si film~\cite{pinheiro2019}. Fig.~\ref{fig_MOI_smooth}(e) shows examples of the spatial profile of the induction component $B_z$ taken for different applied fields, being obtained from an average of 208 $\mu$m wide strip around the colored horizontal dashed lines depicted in panel (a)-(d). They present the typical critical state-like profile for different fields, showing the flux penetration progression for each region. The zigzag patterns, mainly on the right half of the sample, are due to domain boundaries in the MO indicator and have little effect on the flux distribution in the superconducting film. They also become visible as an asymmetry between the peaks on the left and the right edges and in the shielded portion in the flux-free Meissner region up to 38~Oe. Furthermore, the blue line shows a less pronounced penetration due to the thinner local thickness.

\begin{figure}[t]
\centering
\includegraphics[width=1\linewidth]{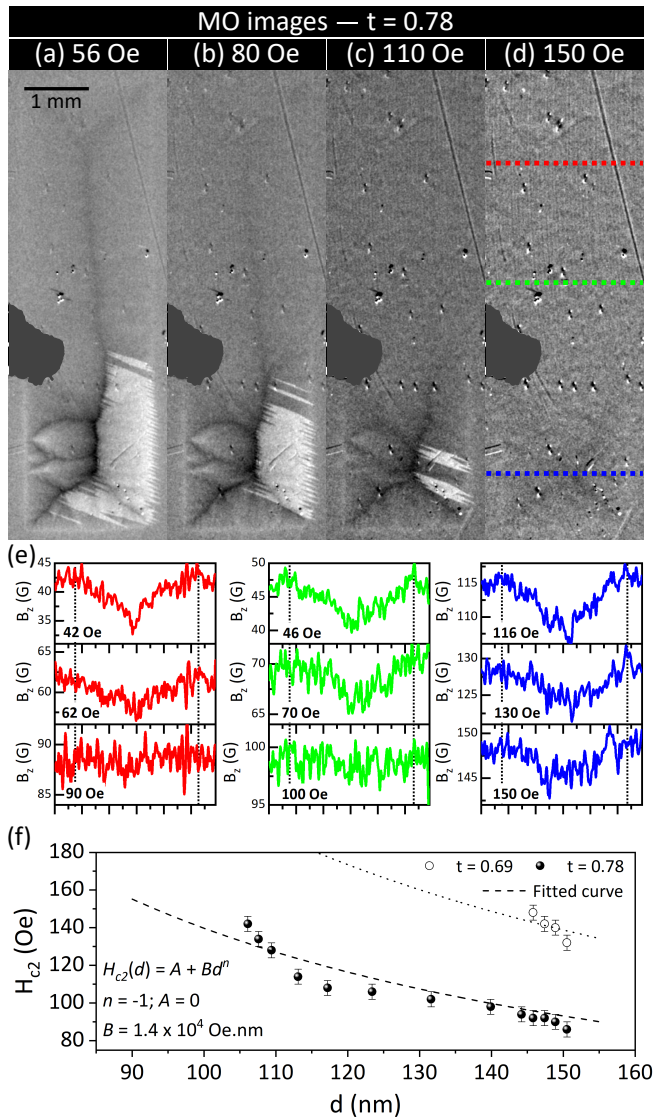}
\caption{MO images of the smooth flux penetration obtained at reduced temperature t~=~0.78 and applied fields of (a) 56~Oe, (b) 80~Oe, (c) 110~Oe, and (d) 150~Oe. Each image was adjusted to optimize the contrast so as to prevent overexposure. (e) Flux density profiles taken along the colored dashed lines depicted in panel (d) for different applied fields. (f) Thickness-dependent upper critical field at reduced temperatures of t~=~0.69 and t~=~0.78 obtained by the analyses of the $B_z$ profile curves.}
\label{fig_Hcxd}
\end{figure}

This penetration pattern is a consequence of the thickness-dependent current density distribution, which causes a higher current density to flow in the thinner region. In order to crudely estimate the critical current density ($J_c$) at different vertical positions, the flux front penetration depth may be evaluated using MO images. Considering the Bean model~\cite{bean1962prl,bean1964rmp} for uniform thin films with stripe geometry, $J_c$ is given by~\cite{brandt1993type,zeldov1994profile}:

\begin{equation} 
\label{Jc-Beanmodel}
    J_{\text{c}} = \frac{\pi H}{d \arccosh\left(\frac{W}{W-a} \right)},
\end{equation}
where $a$ is the flux front penetration depth measured from the long edges. An example of this estimation is given in Fig.~\ref{fig_MOI_smooth}(f) for $t =$ 0.78 and $H =$ 8~Oe. The thickness dependence described by Equation~\eqref{Jc-Beanmodel} arises from geometrical considerations for an infinitely long stripe~\cite{zeldov1994profile}. However, this equation does not explicitly take into account the thickness dependence on $a$. For this reason, we have empirically approximated $J_c$ by an exponentially decaying function of film thickness. This behavior was also reported by Foltyn \textit{et al.} for YBCO~\cite{Foltyn1993} and Huebener and Seher for Pb~\cite{Huebener1969}, but in those cases for uniform films, each with a different thickness. Therefore, decreasing the thickness makes the superconducting film more capable of self-shielding~\cite{Poole,zeldov1994profile}.
The contrast in MO images is a consequence of the penetrated flux and supercurrent density distributions throughout the superconductor~\cite{jooss2002magneto}. Thus, when the superconductor undergoes a phase transition to the normal state, no contrast throughout the image can be recognized. Hence, the critical temperature was determined by MOI in a temperature sweep at zero field after a field cooled (FC) procedure at $H_{FC}$~=~20~Oe. For the wedge-shaped film, the image contrast vanishes homogeneously (not shown) at a single $T_c$, since there is no significant thickness dependence of $T_c$ in the investigated range (90~nm -- 154~nm) for Pb films~\cite{Strongin1970,Ivry2014}. Alternatively, the specimen can also undergo a transition to the normal state above the upper critical field ($H_{c2}$) at a certain temperature. Fig.~\ref{fig_Hcxd}(a)-(d) show MO images taken at $t$~=~0.78 for applied fields higher than the full penetration field throughout the film after a ZFC procedure. One can observe that the thicker region of the specimen fades out when compared to the thinner part. In panel (a), four current domains are separated by two V-shaped d-lines in the extremes and a nearly vertical d-line in the center of the film. At 80~Oe [panel(b)], the borders of the upper part of the sample practically vanish, but the top V-shaped d-line can still be barely discerned together with the vertical one. At a higher field (110~Oe), one can distinguish the persistence of the superconducting state (below the smudge) thanks to the flux distribution and the d-lines.
The evolution of these upper d-lines can be seen in panel (d), taken at 150~Oe, where a X-shaped d-line can still be recognized. These observations demonstrate that the upper critical field depends inversely on the film thickness and reinforces the role played by the thickness variation on the superconducting properties of thin films.

To obtain $H_{c2}(d)$ curves at different temperatures, flux density profiles along the sample width were analyzed for several thicknesses. Fig.~\ref{fig_Hcxd}(e) shows examples of the spatial profile of $B_z$ for different fields, being obtained from an average on a 208 $\mu$m wide strip around the colored horizontal dashed lines depicted in panel (d). At low fields, the Bean-like profile can be recognized for all regions despite restrictions related to the experimental uncertainty that has a standard deviation of 4~G~\cite{davi2021}. As the applied field is increased, the $B_z$ profile becomes flatter, however, a slope towards the center of the sample can still be perceived. When the upper critical field $H_{c2}$ is reached, a roughly position-independent $B_z$ profile takes place as, for instance, in the red and green profiles for 90~Oe and 100~Oe, respectively. By identifying $H_{c2}$ at different thicknesses, a curve of the upper critical field versus thickness can be plotted. Fig.~\ref{fig_Hcxd}(f) shows the $H_{c2}(d)$ taken from the MO images at reduced temperatures of 0.69 and 0.78. In the former (open circles), only the thicker region of the sample undergoes a transition to the normal state. At $t$ = 0.78, $H_{c2}\propto d^{-1}$. A similar behavior was also observed for uniform-thickness Pb films~\cite{HarperTinkham1968,Tinkham1963} and for films made from other metals and alloys~\cite{CodyMiller1968,HarperTinkham1968,Takayama1971,Brandt1971}.

\begin{figure}[t]
\centering
\includegraphics[width=1\linewidth]{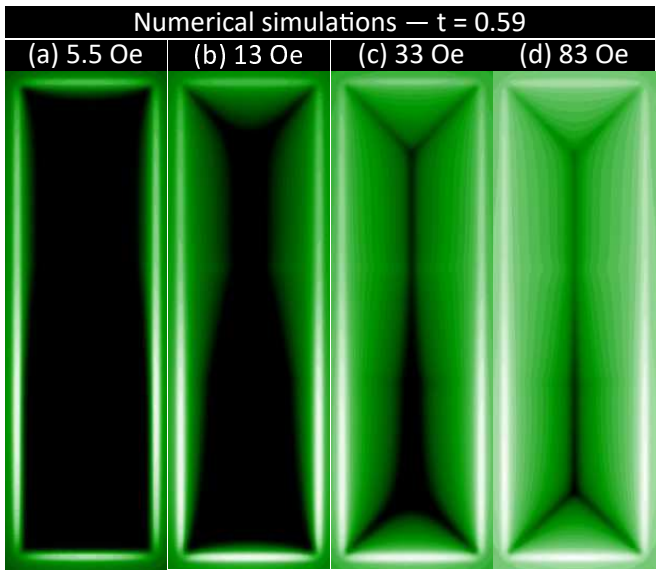}
\caption{Simulated distribution of $B_z$ in a wedge-shaped superconducting film at t = 0.59 after an applied magnetic field from zero to (a) 5.5~Oe, (b) 13~Oe, (c) 33~Oe, and (e) 83~Oe. The image brightness represents the magnitude of $B_z$. The images show a deeper and a less pronounced flux front penetration on the upper region and lower region of the sample, respectively, similar to the MO observations.}
\label{fig_SIM_smooth}
\end{figure}

In order to further analyze the experimental MO results, we have simulated a wedge-shaped superconducting film using the TM model framework~\cite{Denisov2006a,Vestgarden2011}, considering the experimental $J_c(d)$ profile presented in Fig.~\ref{fig_MOI_smooth}(f). Fig.~\ref{fig_SIM_smooth} presents the distribution of magnetic flux density $B_z$ at (a) 5.5~Oe, (b) 13~Oe, (c) 33~Oe, and (d) 83~Oe. The substrate temperature was kept at $t$ = 0.59. In this case, regions where magnetic flux is absent appear as black, whereas maximum field corresponds to the brightest intensity. The main features observed experimentally in Fig.~\ref{fig_MOI_smooth} are captured by the numerical simulations.
In the thicker region on the upper edge, the flux front penetrates deeper and reaches the sample center at around 20~Oe. The flux front in the bottom region, where the supercurrent density is higher, achieves the condition of full penetration at 55~Oe, presenting a better shielding, as in the experiments. 

Another interesting feature is the inhomogeneous brightness along the perimeter of the sample. The lower region of the rim is brighter due to a better shielding ability (higher $J_c$), leading to a higher concentration of magnetic flux at the edges. This increase in the brightness along the bottom rim of the sample is also observed for the MO images, which can be observed in the zoom of the $B_z$ profiles close to the edges in Fig.~\ref{fig_MOI_smooth}(e).


\begin{figure}[t]
\centering
\includegraphics[width=1\linewidth]{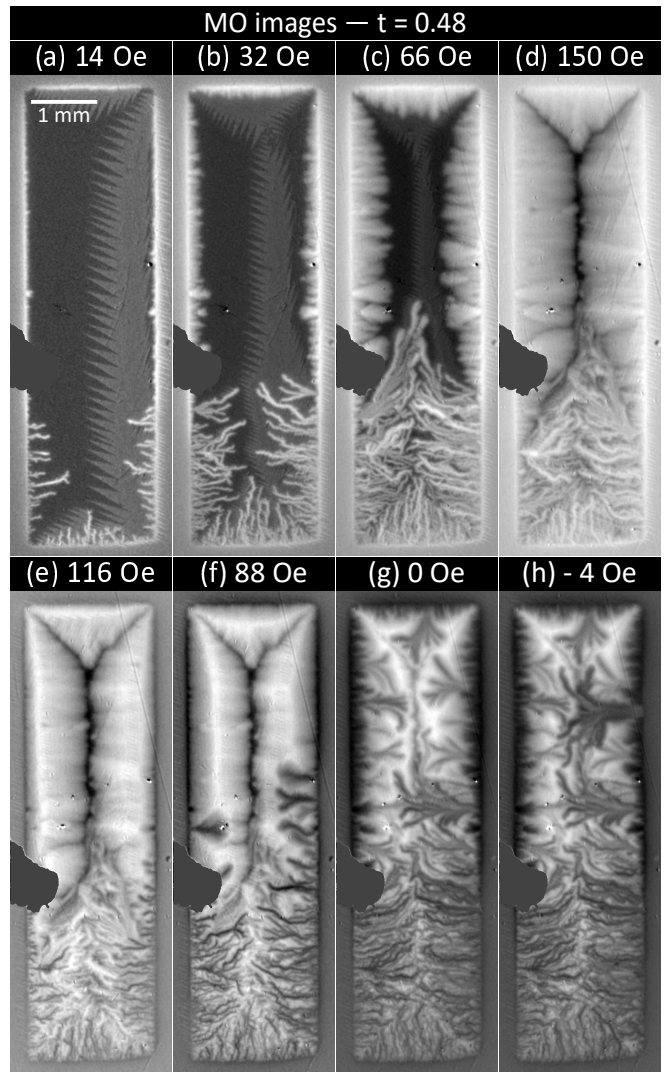}
\caption{MO images of the abrupt penetration of flux avalanches taken at t = 0.48 and fields of (a) 14~Oe, (b) 32~Oe, (c) 66~Oe, and up to (d) 150~Oe, after a ZFC procedure. Then, the field is decreased down to (e) 116~Oe, (f) 88~Oe, (g) 0~Oe (the remnant state), and (h) -4~Oe. Each image was adjusted to optimize the contrast and avoid overexposure.}
\label{fig_MOI_ava}
\end{figure}

Besides the described smooth flux penetration, when the magnetic diffusion is faster than the thermal diffusion, thermomagnetic instabilities may develop and cause flux avalanches~\cite{mints_critical_1981}. This abrupt penetration regime should also be affected by the thickness gradient. Precisely, it corroborates the previous analysis of $J_c$ distribution throughout the specimen, where one expects to trigger more flux avalanches along the edges with higher $J_c$~\cite{Menghini2005,colauto2020controlling}. Fig.~\ref{fig_MOI_ava} shows a series of MO images taken at $t$ = 0.48 after a ZFC procedure and slowly increasing the applied field up to 150~Oe and then decreasing it to negative values. For increasing fields, the penetration pattern is non-uniform as flux avalanches are nucleated in the thinner region only, whereas from the upper thicker part the penetration is smooth. It is somehow similar to the intriguing effect observed by Albrecht \textit{et. al.}~\cite{albrechtprl2007} for a MgB$_2$ film grown onto a vicinal substrate, which results in a current anisotropy of 6\% among the edges.
Numerical simulations using TM model also presented such avalanche preferential propagation~\cite{JingZhou2016} for different current anisotropies. This phenomenon also appears for the wedged-shaped film, which has a difference of around 400\% in $J_c$ between the thinner and thicker regions.

The morphology of the flux avalanches depends strongly on the temperature, as reported by several authors~\cite{Johansen2002,Welling2004,Menghini2005,Vestgarden2011,Vestgarden2013arxiv,Vestgarden2013SUST}. Our results show that even for a fixed temperature, but under different conditions of $H$ and $J_c$ modulated by the thickness variation, the morphology of the avalanches depends on thickness. Fig.~\ref{fig_MOI_ava}(a) shows the first avalanches nucleated exclusively at the thinner region of the sample, where $J_c$ is higher. In this case, the dendrites present many long and thin fingerlike patterns, almost without ramifications. At 32~Oe, branched medium-sized dendrites for thicknesses around 105~nm (2~mm above the bottom edge) and new longer fingers at the lower part emerged. For an applied field of 66~Oe, even larger and more branched dendrites occur whereas the thicker part develops the critical state-like penetration. Another important feature of the avalanches in panel (c) is their size and the bending of their ramifications toward the thicker zone. In other words, when a region of the sample has no avalanches yet, the first branches point towards the center of the sample, as expected. However, avalanches triggered at higher fields close to the thicker zone deviate and propagate towards the flux-free Meissner region. A possible explanation for this effect relies on the fact that avalanches avoid crossing the existing ones, as described earlier by Johansen \textit{et al.}~\cite{Johansen2001}, and there is still flux-free area in the thicker region, while the thinner region is already filled with previously triggered flux avalanches. Similar behavior has also been observed in a work by Choi and coauthors~\cite{Choi2008} in which avalanches bend around a gold rim covering only half of a MgB$_2$ film. Fig.~\ref{fig_MOI_ava}(d) illustrates the coexistence of the smooth penetration and the avalanche regimes in a single sample: at the upper half of the film the flux front already reached the full penetration state whereas avalanches keep on developing in the thinner part. In this case, some branches are supplanted by the Bean-like penetration around the grayish spot. It is also important to mention that avalanches are not triggered in the thicker zone as a consequence of the lateral heat diffusion, which is larger for this portion, suppressing the occurrence of avalanches~\cite{Vestgarden2013SUST} .

\begin{figure}[t]
\centering
\includegraphics[width=1\linewidth]{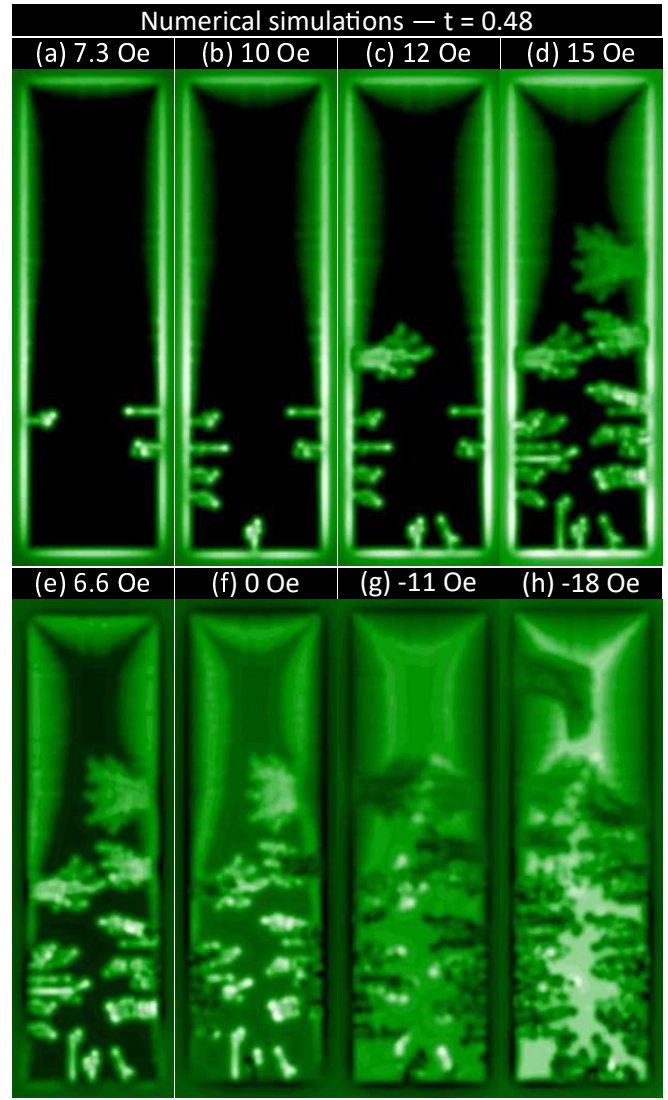}
\caption{Simulated flux penetration into the wedge-shaped film obtained at t = 0.48 for applied fields gradually increasing to (a) 7.3~Oe, (b) 10~Oe, (c) 12~Oe, (d) 15~Oe, after a ZFC procedure. Then, the field is decreased down to (e) 6.6~Oe, (f) the remnant state (0~Oe), (g) -11~Oe, and (h) -18~Oe.}
\label{fig_SIM_ava}
\end{figure}

Additionally, as the critical current density is thickness dependent, the field needed to trigger avalanches increases for increasing thicknesses, as found by Vestg\aa rden and coauthors~\cite{Vestgarden2013arxiv}. Considering the $J_c(d)$ dependency for the wedge-shaped film and that the thicknesses where avalanches occur ranges from 90~nm to 105~nm (a variation of 17\%), an important ingredient for the above described avalanche morphology is the flux front penetration depth before its occurrence at each position $y$ (or thickness)~\cite{Vestgarden2013SUST}. In other words, the deeper initial penetration produces larger and more branching avalanches. 

Contrary to Albrecht \textit{et al.}~\cite{albrechtprl2007} findings in MgB$_2$ films with anisotropic pinning, non-uniform flux penetration is hysteretic for decreasing fields after a maximum of 150~Oe, i.e., negative field-polarity avalanches (or anti-avalanches) are triggered throughout all four sample edges and their morphology change between the thinner and thicker parts. In this case, anti-avalanches nucleate and propagate into the film already populated by positive flux trapped by the pinning centers, still leaving an imprint of positive induction field $B_z$ for $H>0$~\cite{Pinheiro2020}. Panels (e) through (h) of Fig.~\ref{fig_MOI_ava} show anti-avalanches for 116~Oe, 88~Oe, the remnant state, and -4~Oe, respectively. Fingerlike anti-dendrites are triggered along the thinner edges, whereas medium-sized dendrites are nucleated in the middle of the film where only smooth penetration took place for increasing fields. At the remnant state and -4~Oe, several highly branched anti-dendrites can be seen along all edges, even at the top one, showing an interesting hysteresis in the flux penetration for increasing and decreasing applied magnetic fields.

Fig.~\ref{fig_SIM_ava} shows the simulation results for the avalanche regime. At lower fields (7.3~Oe and 10~Oe) after a ZFC procedure, avalanches are triggered in the thinner portion and present a fingerlike shape, whereas at intermediate fields (12~Oe) a branched avalanche is triggered on the left edge. Further avalanches are nucleated at the maximum field of 15~Oe and the Bean-like penetration develops along the upper part of the sample. For decreasing fields, the region out of the sample becomes dark, the field $B_z$ is negative below 6.6 Oe and some small fingerlike anti-dendrites are triggered on the thinner zone at the remnant state.
At negative applied fields, additional anti-dendrites with finger shape are nucleated along the thinner region and branched anti-avalanches are also triggered on the thicker region, one of them where the critical state was established previously. Therefore, the overall features of the flux avalanches were confirmed by numerical simulations.

\begin{figure}[t]
\centering
\includegraphics[width=1\linewidth]{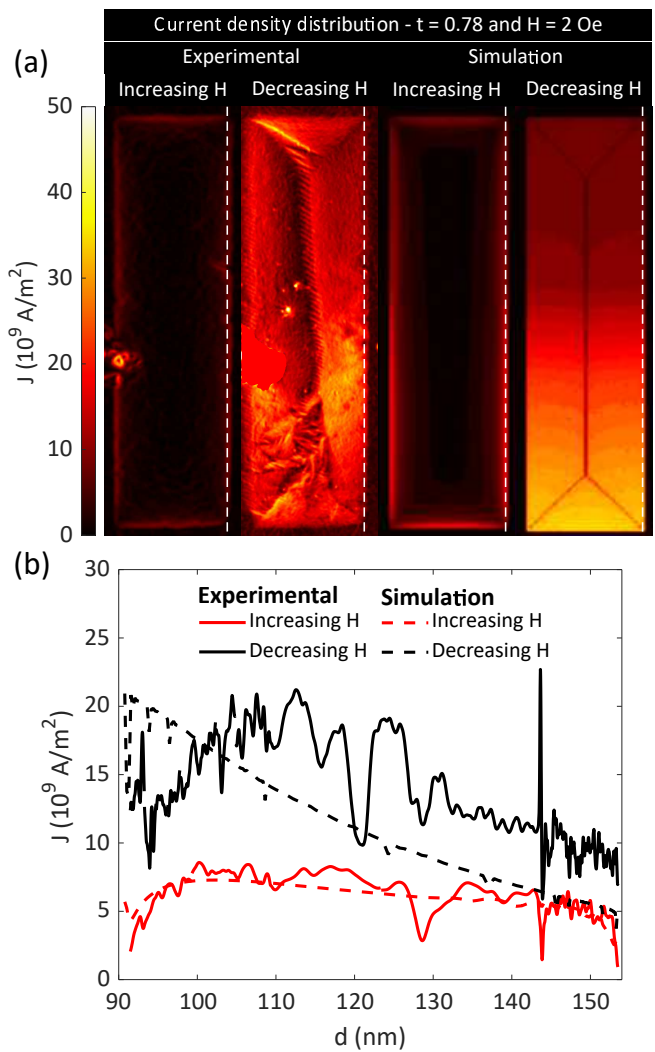}
\caption{Current maps and profiles captured at t = 0.78 for the wedge-shaped film. (a) Experimental and theoretical current maps at 2~Oe, after a ZFC procedure for increasing field, and at the same $H$ after a maximum value of 150~Oe (decreasing field). A solid red smudge at the second panel covers the region where vacuum grease unexpectedly jumped from the cold finger during the experiment. (b) Thickness dependent current for increasing and decreasing fields, after a maximum of 150~Oe, both taken at 2~Oe, along the white vertical dashed line close to the right edge of the film.}
\label{fig_jccomparison}
\end{figure}

The hysteresis between both branches of applied field for triggering flux avalanches can be explained based on the magnetic properties of type-II superconductors together with the TM model. Fig.~\ref{fig_jccomparison} shows current maps and profiles acquired from MO images and simulations for the graded-thickness film. The experimental result was obtained after the transformation from $B_z$ to $J$ maps calculated using the algorithm proposed by Meltzer and coauthors~\cite{meltzer2017reconstruction}, considering the proper thickness profile shown in Fig.~\ref{fig_sample}(c). Moreover, we chose the MO image taken at $t$ = 0.78 and $H$ = 2~Oe to minimize artifacts caused by the presence of zigzag domains in the MO indicator. For increasing fields, the shielding current is mainly confined along the perimeter where the magnetic flux is penetrated and decreases towards the center of the sample, as depicted in Fig.~\ref{fig_jccomparison}(a) for both experimental and simulation data. For decreasing fields after a maximum value high enough to reach the full penetration state, a more complex scenario develops because of the flux penetration and the currents throughout the film.
Although in the experimental MO image for decreasing fields, garnet domains along the flux front and defects on the edges affect the results, the comparison to numerical calculations yields a satisfactory correspondence for both increasing and decreasing fields.



Fig.~\ref{fig_jccomparison}(b) shows $J$ profiles taken along the white vertical dashed line close to the right edge of the sample. The comparison between simulation and experimental data indicate a good agreement despite the large fluctuations observed in the latter. These fluctuations arise from non-uniform flux penetration due to imperfections on the edges. The main observed feature is that both experiment and simulation  show an enhancement of the current density at same $H$ during the decrease of the applied field. It is important to note that the experimental $J$ profiles are obtained directly from the MO images, being not related to the critical state models. This explains the discrepancy between experimental and simulation data for decreasing $H$. For the latter, $J$ is bound by the exponential relationship revealed for $J_c(d)$ in Fig.~\ref{fig_MOI_smooth}(f), as the simulations do not account for an existing $J_c(B)$ dependency.


Based on these $J$ profiles, avalanches are triggered in thicker regions for decreasing fields in part as a consequence of a higher local $J$. An important ingredient in this case, is the positive feedback due to Joule heating, which depends strongly on the critical current density~\cite{Vestgarden2013SUST}. This hysteresis in the threshold fields for the occurrence of flux avalanches was firstly described by Qviller \textit{et al.}~\cite{Qviller2010}. The main idea is related to the difference between the applied field $H$ and the local field $B_z$ close to the edges where avalanches are usually nucleated. Once the field is reduced after a maximum value, the currents close to the edges reverse and $B_z$ reaches a minimum value for a higher value of $H$ in this branch when compared to increasing fields, resulting in higher $J_c(B_z)$ curves for decreasing applied fields.
Therefore, the hysteresis between the increasing and decreasing fields for triggering flux avalanches can be completely understood based on the TM model, as well as the regions where avalanches are triggered and their huge diversity in morphology. 

\section{Conclusions}

In summary, we successfully prepared a rectangular wedge-shaped film of Pb employing shadow-mask evaporation. We observed a strong $J_c(d)$ dependency, as the critical current increased by over a factor of four when going from the thicker (154~nm) to the thinner (90~nm) edge. When the upper critical field is reached for a certain thickness, the superconducting region undergoes a transition to the normal state revealed by a constant value of $B_z$ along the width of the film. Moreover, the variation of $J_c$ with thickness promotes a great diversity of flux avalanche patterns at a fixed temperature. Based on quantitative MOI, the hysteresis of $J$ profiles for increasing and decreasing applied fields explains the occurrence of antiavalanches in the upper half of the specimen. Taking into account the experimental curve $J_c(d)$, the numerical simulations based on the TM model presented a consistent correspondence with the experimental results.

These findings show that fabricating wedge-shaped specimens, i.e., exhibiting a thickness gradient, is an interesting strategy to investigate in one and the same sample the material properties dependent on thickness. For superconductors, $T_c$ as well as pinning properties can be modulated with suitable thickness. 



\section*{Acknowledgements}

This work was partially supported by the São Paulo Research Foundation (FAPESP, Grant 2021/08781-8), the National Council for Scientific and Technological Development (CNPq, Grants 433700/2018-1, 431974/2018-7, 316602/2021-3, and 309928/2018-4), and Coordenação de Aperfeiçoamento de Pessoal de Nível Superior – Brasil (CAPES) - Finance Code 001. C. X. and L. J. acknowledge the support by the National Natural Science Foundation of China (Grants No. 11972298). L. J. was supported by the China Scholarship Council.
The authors would like to thank Laboratory of Structural Characterization (LCE/DEMa/UFSCar) for grating use of its facilities.

L.B.L.G.P. and L.J. contributed equally to this work.

\appendix

\section{Wedge-film deposition}
\label{sec:SampleDeposition}

A new all-metal high vacuum chamber was developed to deposit Pb films employing the conventional thermal evaporation technique~\cite{smith1995thin} to achieve controllable conditions to grow low melting point metals. Pb films were fabricated by evaporating 99.999\% pure lead placed in a tungsten boat onto a Si (100) substrate. A base pressure of 8 x $10^{-7}$ torr is achieced using a liquid-nitrogen cold trap far from the steady substrate holder, which was kept at room temperature. Due to the quick degradation and short lifetime of Pb when exposed to air, a cap layer of Ge was deposited to protect it against oxidation. Throughout the deposition, a quartz crystal microbalance (QCM) based on the OpenQCM project~\cite{openQCM} was used to monitor the film thickness during deposition. 

To avoid contamination during the film deposition, target metals and substrates were cleaned by chemical wet etching. We prepared the following solutions: HF:HNO$_3$ with volume ratio of 1:1 for Ge; CH$_3$COOH:H$_2$O$_2$ with ratio 1:1 for Pb; and Piranha etch for the Si wafers~\cite{metaletchants1990}.

\begin{figure}[b]
\centering
\includegraphics[width=1\linewidth]{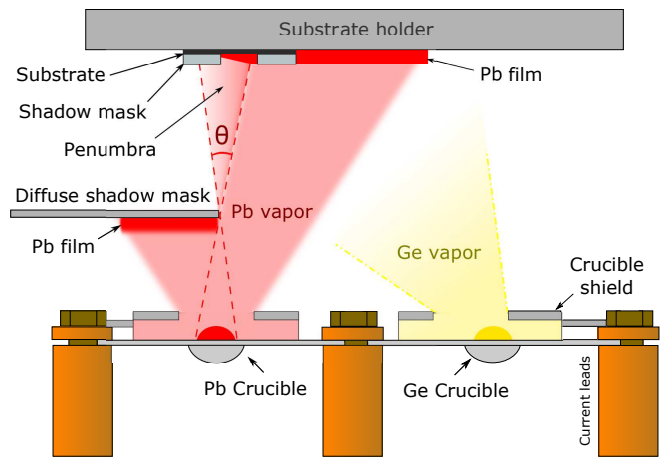}
\caption{Schematic diagram of the interior of the thermal evaporator chamber. There is a diffuse-shadow mask between the Pb crucible and the substrate holder to produce a penumbra region where a film with thickness gradient is grown. This image is out of scale.}
\label{fig_deposition}
\end{figure}

In order to produce a wedge-shaped thin film, a Diffuse-Shadow Mask (DSM) was built and placed inside the chamber between the Pb crucible and the substrate holder, as illustrated in Fig.~\ref{fig_deposition}. On the left, the Pb vapor is partially blocked by the crucible shield. The DSM is placed partially covering the Pb crucible, which takes advantage of the non-punctual vapor source to create an area of penumbra identified by $\theta$ and delimited by the dashed red lines delineating the Pb atom paths traced from the extremes of the crucible dip to the substrate holder. Red color indicates the uniform and the wedge-shaped deposition regions for Pb, just below the substrate holder as well as below the DSM. The position of the crucibles in relation to the shields and shutters was made to enable the protective Ge vapor to cover all the wedge-shaped deposition zone. By varying the distance between the DSM and the substrate holder, we are capable of controlling the area of penumbra, i.e., the wedge-shaped deposition zone, and thus change the thickness gradient. 



To delineate the edges, another shadow mask was directly clamped to the substrate. Thus, the resulting film has a rectangular shape of area $2\times 7$~mm$^{2}$ and the thickness gradient along the major length.

\section{SEM/EDS characterization}
\label{sec:EDS}

\begin{figure}[t]
\centering
\includegraphics[width=1\linewidth]{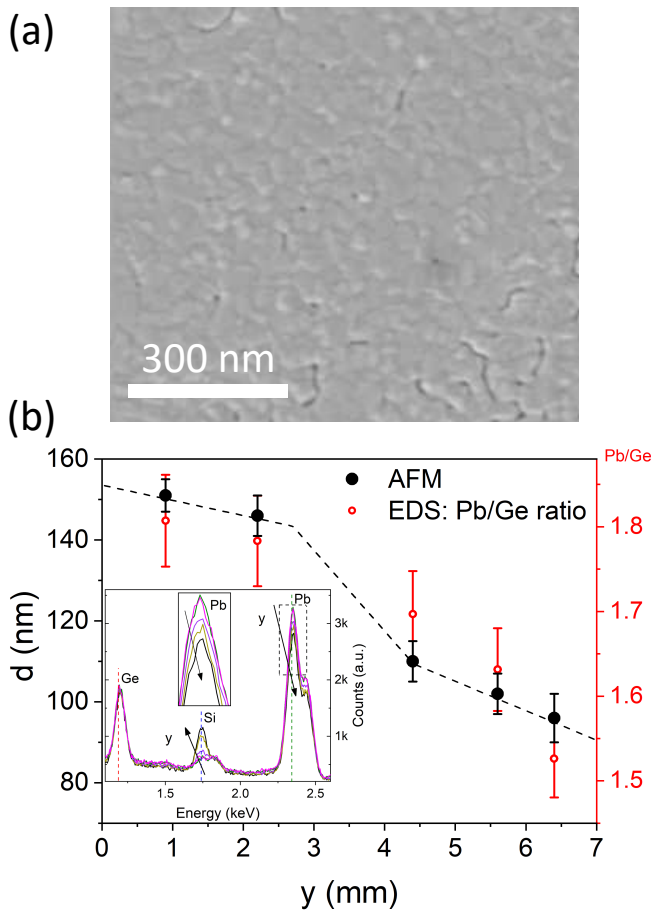}
\caption{(a) Secondary electron SEM image of a co-evaporated sample. (b) Intensity ratio of Pb/Ge is plotted in red in the main panel with its ordinate axis on the right. The inset presents EDS spectra obtained from the regions identified in (a). The thickness versus y position obtained from the AFM is also plotted for comparison.}
\label{fig_EDS}
\end{figure}

It is noteworthy that the surface roughness of the Ge layer (the one with which the AFM tip interacts) is uniform all over the sample, showing grains in the nanometric scale, with an average area around 1100~nm$^2$. One can observe this morphology in the secondary electron SEM image in Fig.~\ref{fig_EDS}(a), which is somewhat similar to the single Pb layers deposited on different conditions shown in Refs.~\citenum{Perrone2013Pb,Lorusso2015Pb,Broitman2016Pb}.

The elemental composition of a given material can be obtained from its EDS spectra~\cite{Reimer1985}. As the detected energy intensity depends on the initial electron beam energy and the average volume of the droplet-shaped interaction region, EDS may also be employed to estimate the film thickness~\cite{Pascual1990edsthick,Ng2006edsthick,Zhuang2009edsthick,Habiger1992,pinheiro2019}.
Most authors use the relation $I_f/I_s \sim d$, where $I_s$ ($I_f$) is the peak intensity of the substrate (film), to obtain the thickness. One way to determine $d$ is to scan the incident electron beam energy to identify at which value the generation of X-rays is confined only to the sample film~\cite{Pascual1990edsthick}. Then a calibrated curve can be generated from known thicknesses to calculate the thickness of unknown samples. In our case, we kept the acceleration voltage constant at 5~kV, which is large enough to interact with all different layers, including the substrate, and use the Pb/Ge intensity ratio to map the relative profile of the Pb layer along the gradient thickness of the film.

Thus, we chose the same positions where the AFM imaging was performed and monitored the relative heights of the X-ray peaks of the three main elements, assuming that the initial energy of the electron beam penetrates deeply enough to interact with all layers concomitantly. The inset in Fig.~\ref{fig_EDS}(b) shows the five spectra for the different regions indicated in Fig.~\ref{fig_sample}(a) in the range of 1.08 to 2.60 keV, where one can identify the Ge-L$\alpha$ (1.188 keV), Si-K$\alpha$ (1.739 keV), and Pb-M$\alpha$ (2.342 keV) characteristic radiation~\cite{Goldstein2003_analy,Reed2005m_analy}. Since the Ge layer is uniform, its peak height is constant for all vertical positions. As expected, the peak height related to Pb decreases as the thickness decreases, as indicated in the zoom-up in the center of the graph. Conversely, a higher signal due to Si is obtained as the superconducting film becomes thinner as an indication that the electron beam interacts more with the substrate. 
Black arrows were added to the peaks to indicate the decrease in thickness. 
In contrast to the previous works cited, we use the intensity peak ratio between Pb/Ge, which is directly proportional to $d$, to represent the relative thickness variation as depicted in the main panel of Fig.~\ref{fig_EDS}(b). The error bars were taken as 5\% for the Pb/Ge ratio based on the relative error of the element amounts. One can observe that the data follows the general behavior of the thickness determined by AFM despite the different scale. Therefore, this data confirms the general behavior of the non-linear curve for the thickness variation.

\bibliography{references.bib}

\begin{thebibliography}{104}%
\makeatletter
\providecommand \@ifxundefined [1]{%
 \@ifx{#1\undefined}
}%
\providecommand \@ifnum [1]{%
 \ifnum #1\expandafter \@firstoftwo
 \else \expandafter \@secondoftwo
 \fi
}%
\providecommand \@ifx [1]{%
 \ifx #1\expandafter \@firstoftwo
 \else \expandafter \@secondoftwo
 \fi
}%
\providecommand \natexlab [1]{#1}%
\providecommand \enquote  [1]{``#1''}%
\providecommand \bibnamefont  [1]{#1}%
\providecommand \bibfnamefont [1]{#1}%
\providecommand \citenamefont [1]{#1}%
\providecommand \href@noop [0]{\@secondoftwo}%
\providecommand \href [0]{\begingroup \@sanitize@url \@href}%
\providecommand \@href[1]{\@@startlink{#1}\@@href}%
\providecommand \@@href[1]{\endgroup#1\@@endlink}%
\providecommand \@sanitize@url [0]{\catcode `\\12\catcode `\$12\catcode
  `\&12\catcode `\#12\catcode `\^12\catcode `\_12\catcode `\%12\relax}%
\providecommand \@@startlink[1]{}%
\providecommand \@@endlink[0]{}%
\providecommand \url  [0]{\begingroup\@sanitize@url \@url }%
\providecommand \@url [1]{\endgroup\@href {#1}{\urlprefix }}%
\providecommand \urlprefix  [0]{URL }%
\providecommand \Eprint [0]{\href }%
\providecommand \doibase [0]{http://dx.doi.org/}%
\providecommand \selectlanguage [0]{\@gobble}%
\providecommand \bibinfo  [0]{\@secondoftwo}%
\providecommand \bibfield  [0]{\@secondoftwo}%
\providecommand \translation [1]{[#1]}%
\providecommand \BibitemOpen [0]{}%
\providecommand \bibitemStop [0]{}%
\providecommand \bibitemNoStop [0]{.\EOS\space}%
\providecommand \EOS [0]{\spacefactor3000\relax}%
\providecommand \BibitemShut  [1]{\csname bibitem#1\endcsname}%
\let\auto@bib@innerbib\@empty
\bibitem [{\citenamefont {Wendin}(2017)}]{Wendin2017}%
  \BibitemOpen
  \bibfield  {author} {\bibinfo {author} {\bibfnamefont {G.}~\bibnamefont
  {Wendin}},\ }\href@noop {} {\bibfield  {journal} {\bibinfo  {journal} {Rep.
  Prog. Phys.}\ }\textbf {\bibinfo {volume} {80}} (\bibinfo {year}
  {2017})}\BibitemShut {NoStop}%
\bibitem [{\citenamefont {Kjaergaard}\ \emph {et~al.}(2020)\citenamefont
  {Kjaergaard}, \citenamefont {Schwartz}, \citenamefont {Braum{\"u}ller},
  \citenamefont {Krantz}, \citenamefont {Wang}, \citenamefont {Gustavsson},\
  and\ \citenamefont {Oliver}}]{kjaergaard2020superconducting}%
  \BibitemOpen
  \bibfield  {author} {\bibinfo {author} {\bibfnamefont {M.}~\bibnamefont
  {Kjaergaard}}, \bibinfo {author} {\bibfnamefont {M.~E.}\ \bibnamefont
  {Schwartz}}, \bibinfo {author} {\bibfnamefont {J.}~\bibnamefont
  {Braum{\"u}ller}}, \bibinfo {author} {\bibfnamefont {P.}~\bibnamefont
  {Krantz}}, \bibinfo {author} {\bibfnamefont {J.~I.-J.}\ \bibnamefont {Wang}},
  \bibinfo {author} {\bibfnamefont {S.}~\bibnamefont {Gustavsson}}, \ and\
  \bibinfo {author} {\bibfnamefont {W.~D.}\ \bibnamefont {Oliver}},\
  }\href@noop {} {\bibfield  {journal} {\bibinfo  {journal} {Annu. Rev.
  Condens. Matter Phys.}\ }\textbf {\bibinfo {volume} {11}},\ \bibinfo {pages}
  {369} (\bibinfo {year} {2020})}\BibitemShut {NoStop}%
\bibitem [{\citenamefont {Hadfield}(2009)}]{Hadfield2009}%
  \BibitemOpen
  \bibfield  {author} {\bibinfo {author} {\bibfnamefont {R.~H.}\ \bibnamefont
  {Hadfield}},\ }\href {\doibase 10.1038/nphoton.2009.230} {\bibfield
  {journal} {\bibinfo  {journal} {Nat. Photonics}\ }\textbf {\bibinfo {volume}
  {3}},\ \bibinfo {pages} {696} (\bibinfo {year} {2009})}\BibitemShut {NoStop}%
\bibitem [{\citenamefont {Natarajan}\ \emph {et~al.}(2012)\citenamefont
  {Natarajan}, \citenamefont {Tanner},\ and\ \citenamefont
  {Hadfield}}]{Natarajan2012}%
  \BibitemOpen
  \bibfield  {author} {\bibinfo {author} {\bibfnamefont {C.~M.}\ \bibnamefont
  {Natarajan}}, \bibinfo {author} {\bibfnamefont {M.~G.}\ \bibnamefont
  {Tanner}}, \ and\ \bibinfo {author} {\bibfnamefont {R.~H.}\ \bibnamefont
  {Hadfield}},\ }\href {\doibase 10.1088/0953-2048/25/6/063001} {\bibfield
  {journal} {\bibinfo  {journal} {Supercond. Sci. Technol.}\ }\textbf {\bibinfo
  {volume} {25}},\ \bibinfo {pages} {063001} (\bibinfo {year}
  {2012})}\BibitemShut {NoStop}%
\bibitem [{\citenamefont {Franssila}(2010)}]{SamiFranssila2010}%
  \BibitemOpen
  \bibfield  {author} {\bibinfo {author} {\bibfnamefont {S.}~\bibnamefont
  {Franssila}},\ }\href
  {http://onlinelibrary.wiley.com/book/10.1002/9781119990413} {\emph {\bibinfo
  {title} {Introduction to Microfabrication}}},\ \bibinfo {edition} {2nd}\ ed.\
  (\bibinfo  {publisher} {Wiley Online Library},\ \bibinfo {year} {2010})\ pp.\
  \bibinfo {pages} {1--518}\BibitemShut {NoStop}%
\bibitem [{\citenamefont {Luna-Moreno}\ and\ \citenamefont
  {Monzón-Hernández}(2007)}]{Luna2007unif}%
  \BibitemOpen
  \bibfield  {author} {\bibinfo {author} {\bibfnamefont {D.}~\bibnamefont
  {Luna-Moreno}}\ and\ \bibinfo {author} {\bibfnamefont {D.}~\bibnamefont
  {Monzón-Hernández}},\ }\href {\doibase
  https://doi.org/10.1016/j.apsusc.2007.04.059} {\bibfield  {journal} {\bibinfo
   {journal} {Appl. Surf. Sci.}\ }\textbf {\bibinfo {volume} {253}},\ \bibinfo
  {pages} {8615} (\bibinfo {year} {2007})}\BibitemShut {NoStop}%
\bibitem [{\citenamefont {Yamamura}\ \emph {et~al.}(2008)\citenamefont
  {Yamamura}, \citenamefont {Shimada},\ and\ \citenamefont
  {Mori}}]{Yamamura2008unif}%
  \BibitemOpen
  \bibfield  {author} {\bibinfo {author} {\bibfnamefont {K.}~\bibnamefont
  {Yamamura}}, \bibinfo {author} {\bibfnamefont {S.}~\bibnamefont {Shimada}}, \
  and\ \bibinfo {author} {\bibfnamefont {Y.}~\bibnamefont {Mori}},\ }\href
  {\doibase https://doi.org/10.1016/j.cirp.2008.03.132} {\bibfield  {journal}
  {\bibinfo  {journal} {CIRP Annals}\ }\textbf {\bibinfo {volume} {57}},\
  \bibinfo {pages} {567} (\bibinfo {year} {2008})}\BibitemShut {NoStop}%
\bibitem [{\citenamefont {Choi}\ \emph {et~al.}(2010)\citenamefont {Choi},
  \citenamefont {Kim}, \citenamefont {Park},\ and\ \citenamefont
  {Lee}}]{Choi2010unif}%
  \BibitemOpen
  \bibfield  {author} {\bibinfo {author} {\bibfnamefont {Y.-O.}\ \bibnamefont
  {Choi}}, \bibinfo {author} {\bibfnamefont {N.-H.}\ \bibnamefont {Kim}},
  \bibinfo {author} {\bibfnamefont {J.-S.}\ \bibnamefont {Park}}, \ and\
  \bibinfo {author} {\bibfnamefont {W.-S.}\ \bibnamefont {Lee}},\ }\href
  {\doibase https://doi.org/10.1016/j.mseb.2010.03.072} {\bibfield  {journal}
  {\bibinfo  {journal} {Mater. Sci. Eng. B}\ }\textbf {\bibinfo {volume}
  {171}},\ \bibinfo {pages} {73} (\bibinfo {year} {2010})}\BibitemShut
  {NoStop}%
\bibitem [{\citenamefont {Wang}\ \emph {et~al.}(2018)\citenamefont {Wang},
  \citenamefont {Fu}, \citenamefont {Song}, \citenamefont {Chu}, \citenamefont
  {Gibson}, \citenamefont {Li}, \citenamefont {Shi},\ and\ \citenamefont
  {Wu}}]{Wang2018unif}%
  \BibitemOpen
  \bibfield  {author} {\bibinfo {author} {\bibfnamefont {B.}~\bibnamefont
  {Wang}}, \bibinfo {author} {\bibfnamefont {X.}~\bibnamefont {Fu}}, \bibinfo
  {author} {\bibfnamefont {S.}~\bibnamefont {Song}}, \bibinfo {author}
  {\bibfnamefont {H.~O.}\ \bibnamefont {Chu}}, \bibinfo {author} {\bibfnamefont
  {D.}~\bibnamefont {Gibson}}, \bibinfo {author} {\bibfnamefont
  {C.}~\bibnamefont {Li}}, \bibinfo {author} {\bibfnamefont {Y.}~\bibnamefont
  {Shi}}, \ and\ \bibinfo {author} {\bibfnamefont {Z.}~\bibnamefont {Wu}},\
  }\href {https://www.mdpi.com/2079-6412/8/9/325} {\bibfield  {journal}
  {\bibinfo  {journal} {Coatings}\ }\textbf {\bibinfo {volume} {8}} (\bibinfo
  {year} {2018})}\BibitemShut {NoStop}%
\bibitem [{\citenamefont {Kim}\ \emph {et~al.}(2018)\citenamefont {Kim},
  \citenamefont {Hu}, \citenamefont {Nam}, \citenamefont {Kim},\ and\
  \citenamefont {Yoo}}]{Kim2018unif}%
  \BibitemOpen
  \bibfield  {author} {\bibinfo {author} {\bibfnamefont {S.-G.}\ \bibnamefont
  {Kim}}, \bibinfo {author} {\bibfnamefont {Q.}~\bibnamefont {Hu}}, \bibinfo
  {author} {\bibfnamefont {K.-B.}\ \bibnamefont {Nam}}, \bibinfo {author}
  {\bibfnamefont {M.~J.}\ \bibnamefont {Kim}}, \ and\ \bibinfo {author}
  {\bibfnamefont {J.-B.}\ \bibnamefont {Yoo}},\ }\href {\doibase
  https://doi.org/10.1016/j.cplett.2018.03.013} {\bibfield  {journal} {\bibinfo
   {journal} {Chem. Phys. Lett.}\ }\textbf {\bibinfo {volume} {698}},\ \bibinfo
  {pages} {157} (\bibinfo {year} {2018})}\BibitemShut {NoStop}%
\bibitem [{\citenamefont {Knehr}\ \emph {et~al.}(2021)\citenamefont {Knehr},
  \citenamefont {Ziegler}, \citenamefont {Linzen}, \citenamefont {Ilin},
  \citenamefont {Schanz}, \citenamefont {Plentz}, \citenamefont {Diegel},
  \citenamefont {Schmidt}, \citenamefont {Il’ichev},\ and\ \citenamefont
  {Siegel}}]{Knehr2021unif}%
  \BibitemOpen
  \bibfield  {author} {\bibinfo {author} {\bibfnamefont {E.}~\bibnamefont
  {Knehr}}, \bibinfo {author} {\bibfnamefont {M.}~\bibnamefont {Ziegler}},
  \bibinfo {author} {\bibfnamefont {S.}~\bibnamefont {Linzen}}, \bibinfo
  {author} {\bibfnamefont {K.}~\bibnamefont {Ilin}}, \bibinfo {author}
  {\bibfnamefont {P.}~\bibnamefont {Schanz}}, \bibinfo {author} {\bibfnamefont
  {J.}~\bibnamefont {Plentz}}, \bibinfo {author} {\bibfnamefont
  {M.}~\bibnamefont {Diegel}}, \bibinfo {author} {\bibfnamefont
  {H.}~\bibnamefont {Schmidt}}, \bibinfo {author} {\bibfnamefont
  {E.}~\bibnamefont {Il’ichev}}, \ and\ \bibinfo {author} {\bibfnamefont
  {M.}~\bibnamefont {Siegel}},\ }\href {\doibase 10.1116/6.0001126} {\bibfield
  {journal} {\bibinfo  {journal} {J. Vac. Sci. Technol. A}\ }\textbf {\bibinfo
  {volume} {39}},\ \bibinfo {pages} {052401} (\bibinfo {year}
  {2021})}\BibitemShut {NoStop}%
\bibitem [{\citenamefont {Zarnetta}\ \emph {et~al.}(2010)\citenamefont
  {Zarnetta}, \citenamefont {Takahashi}, \citenamefont {Young}, \citenamefont
  {Savan}, \citenamefont {Furuya}, \citenamefont {Thienhaus}, \citenamefont
  {Maaß}, \citenamefont {Rahim}, \citenamefont {Frenzel}, \citenamefont
  {Brunken}, \citenamefont {Chu}, \citenamefont {Srivastava}, \citenamefont
  {James}, \citenamefont {Takeuchi}, \citenamefont {Eggeler},\ and\
  \citenamefont {Ludwig}}]{Zarnetta2010}%
  \BibitemOpen
  \bibfield  {author} {\bibinfo {author} {\bibfnamefont {R.}~\bibnamefont
  {Zarnetta}}, \bibinfo {author} {\bibfnamefont {R.}~\bibnamefont {Takahashi}},
  \bibinfo {author} {\bibfnamefont {M.~L.}\ \bibnamefont {Young}}, \bibinfo
  {author} {\bibfnamefont {A.}~\bibnamefont {Savan}}, \bibinfo {author}
  {\bibfnamefont {Y.}~\bibnamefont {Furuya}}, \bibinfo {author} {\bibfnamefont
  {S.}~\bibnamefont {Thienhaus}}, \bibinfo {author} {\bibfnamefont
  {B.}~\bibnamefont {Maaß}}, \bibinfo {author} {\bibfnamefont
  {M.}~\bibnamefont {Rahim}}, \bibinfo {author} {\bibfnamefont
  {J.}~\bibnamefont {Frenzel}}, \bibinfo {author} {\bibfnamefont
  {H.}~\bibnamefont {Brunken}}, \bibinfo {author} {\bibfnamefont {Y.~S.}\
  \bibnamefont {Chu}}, \bibinfo {author} {\bibfnamefont {V.}~\bibnamefont
  {Srivastava}}, \bibinfo {author} {\bibfnamefont {R.~D.}\ \bibnamefont
  {James}}, \bibinfo {author} {\bibfnamefont {I.}~\bibnamefont {Takeuchi}},
  \bibinfo {author} {\bibfnamefont {G.}~\bibnamefont {Eggeler}}, \ and\
  \bibinfo {author} {\bibfnamefont {A.}~\bibnamefont {Ludwig}},\ }\href
  {\doibase https://doi.org/10.1002/adfm.200902336} {\bibfield  {journal}
  {\bibinfo  {journal} {Adv. Funct. Mater.}\ }\textbf {\bibinfo {volume}
  {20}},\ \bibinfo {pages} {1917} (\bibinfo {year} {2010})}\BibitemShut
  {NoStop}%
\bibitem [{\citenamefont {Ludwig}(2019)}]{Ludwig2019}%
  \BibitemOpen
  \bibfield  {author} {\bibinfo {author} {\bibfnamefont {A.}~\bibnamefont
  {Ludwig}},\ }\href {http://dx.doi.org/10.1038/s41524-019-0205-0} {\bibfield
  {journal} {\bibinfo  {journal} {npj Comput. Mater.}\ }\textbf {\bibinfo
  {volume} {5}},\ \bibinfo {pages} {70} (\bibinfo {year} {2019})}\BibitemShut
  {NoStop}%
\bibitem [{\citenamefont {Yuan}\ \emph {et~al.}(2022)\citenamefont {Yuan},
  \citenamefont {Chen}, \citenamefont {Jiang}, \citenamefont {Feng},
  \citenamefont {Lin}, \citenamefont {Yu}, \citenamefont {He}, \citenamefont
  {Zhang}, \citenamefont {Jiang}, \citenamefont {Zhang}, \citenamefont {Shi},
  \citenamefont {Zhang}, \citenamefont {Qin}, \citenamefont {Cheng},
  \citenamefont {Tamura}, \citenamefont {feng Yang}, \citenamefont {Xiang},
  \citenamefont {Hu}, \citenamefont {Takeuchi}, \citenamefont {Jin},\ and\
  \citenamefont {Zhao}}]{Yuan2022}%
  \BibitemOpen
  \bibfield  {author} {\bibinfo {author} {\bibfnamefont {J.}~\bibnamefont
  {Yuan}}, \bibinfo {author} {\bibfnamefont {Q.}~\bibnamefont {Chen}}, \bibinfo
  {author} {\bibfnamefont {K.}~\bibnamefont {Jiang}}, \bibinfo {author}
  {\bibfnamefont {Z.}~\bibnamefont {Feng}}, \bibinfo {author} {\bibfnamefont
  {Z.}~\bibnamefont {Lin}}, \bibinfo {author} {\bibfnamefont {H.}~\bibnamefont
  {Yu}}, \bibinfo {author} {\bibfnamefont {G.}~\bibnamefont {He}}, \bibinfo
  {author} {\bibfnamefont {J.}~\bibnamefont {Zhang}}, \bibinfo {author}
  {\bibfnamefont {X.}~\bibnamefont {Jiang}}, \bibinfo {author} {\bibfnamefont
  {X.}~\bibnamefont {Zhang}}, \bibinfo {author} {\bibfnamefont
  {Y.}~\bibnamefont {Shi}}, \bibinfo {author} {\bibfnamefont {Y.}~\bibnamefont
  {Zhang}}, \bibinfo {author} {\bibfnamefont {M.}~\bibnamefont {Qin}}, \bibinfo
  {author} {\bibfnamefont {Z.~G.}\ \bibnamefont {Cheng}}, \bibinfo {author}
  {\bibfnamefont {N.}~\bibnamefont {Tamura}}, \bibinfo {author} {\bibfnamefont
  {Y.}~\bibnamefont {feng Yang}}, \bibinfo {author} {\bibfnamefont
  {T.}~\bibnamefont {Xiang}}, \bibinfo {author} {\bibfnamefont
  {J.}~\bibnamefont {Hu}}, \bibinfo {author} {\bibfnamefont {I.}~\bibnamefont
  {Takeuchi}}, \bibinfo {author} {\bibfnamefont {K.}~\bibnamefont {Jin}}, \
  and\ \bibinfo {author} {\bibfnamefont {Z.}~\bibnamefont {Zhao}},\ }\href
  {\doibase 10.1038/s41586-021-04305-5} {\bibfield  {journal} {\bibinfo
  {journal} {Nature}\ }\textbf {\bibinfo {volume} {602}},\ \bibinfo {pages}
  {431} (\bibinfo {year} {2022})}\BibitemShut {NoStop}%
\bibitem [{\citenamefont {Palmisano}\ \emph {et~al.}(2010)\citenamefont
  {Palmisano}, \citenamefont {Filippi}, \citenamefont {Baldi}, \citenamefont
  {Slaman}, \citenamefont {Schreuders},\ and\ \citenamefont
  {Dam}}]{Palmisano2010}%
  \BibitemOpen
  \bibfield  {author} {\bibinfo {author} {\bibfnamefont {V.}~\bibnamefont
  {Palmisano}}, \bibinfo {author} {\bibfnamefont {M.}~\bibnamefont {Filippi}},
  \bibinfo {author} {\bibfnamefont {A.}~\bibnamefont {Baldi}}, \bibinfo
  {author} {\bibfnamefont {M.}~\bibnamefont {Slaman}}, \bibinfo {author}
  {\bibfnamefont {H.}~\bibnamefont {Schreuders}}, \ and\ \bibinfo {author}
  {\bibfnamefont {B.}~\bibnamefont {Dam}},\ }\href {\doibase
  10.1016/j.ijhydene.2010.09.001} {\bibfield  {journal} {\bibinfo  {journal}
  {Int. J. Hydrogen Energy}\ }\textbf {\bibinfo {volume} {35}},\ \bibinfo
  {pages} {12574} (\bibinfo {year} {2010})}\BibitemShut {NoStop}%
\bibitem [{\citenamefont {Hiromasa}\ \emph {et~al.}(2021)\citenamefont
  {Hiromasa}, \citenamefont {Takahiro}, \citenamefont {Takumi},\ and\
  \citenamefont {Shogo}}]{Shimizu2021}%
  \BibitemOpen
  \bibfield  {author} {\bibinfo {author} {\bibfnamefont {S.}~\bibnamefont
  {Hiromasa}}, \bibinfo {author} {\bibfnamefont {O.}~\bibnamefont {Takahiro}},
  \bibinfo {author} {\bibfnamefont {M.}~\bibnamefont {Takumi}}, \ and\ \bibinfo
  {author} {\bibfnamefont {S.}~\bibnamefont {Shogo}},\ }\href {\doibase
  10.3389/fnano.2021.724528} {\bibfield  {journal} {\bibinfo  {journal}
  {Frontiers in Nanotechnology}\ }\textbf {\bibinfo {volume} {3}},\ \bibinfo
  {pages} {70} (\bibinfo {year} {2021})}\BibitemShut {NoStop}%
\bibitem [{\citenamefont {Born}\ \emph {et~al.}(2006)\citenamefont {Born},
  \citenamefont {Siegel}, \citenamefont {Hollmann}, \citenamefont {Braak},
  \citenamefont {Golubov}, \citenamefont {Gusakova},\ and\ \citenamefont
  {Kupriyanov}}]{Born2006}%
  \BibitemOpen
  \bibfield  {author} {\bibinfo {author} {\bibfnamefont {F.}~\bibnamefont
  {Born}}, \bibinfo {author} {\bibfnamefont {M.}~\bibnamefont {Siegel}},
  \bibinfo {author} {\bibfnamefont {E.~K.}\ \bibnamefont {Hollmann}}, \bibinfo
  {author} {\bibfnamefont {H.}~\bibnamefont {Braak}}, \bibinfo {author}
  {\bibfnamefont {A.~A.}\ \bibnamefont {Golubov}}, \bibinfo {author}
  {\bibfnamefont {D.~Y.}\ \bibnamefont {Gusakova}}, \ and\ \bibinfo {author}
  {\bibfnamefont {M.~Y.}\ \bibnamefont {Kupriyanov}},\ }\href {\doibase
  10.1103/PhysRevB.74.140501} {\bibfield  {journal} {\bibinfo  {journal} {Phys.
  Rev. B}\ }\textbf {\bibinfo {volume} {74}},\ \bibinfo {pages} {140501}
  (\bibinfo {year} {2006})}\BibitemShut {NoStop}%
\bibitem [{\citenamefont {Antropov}\ \emph {et~al.}(2013)\citenamefont
  {Antropov}, \citenamefont {Kalenkov}, \citenamefont {Kehrle}, \citenamefont
  {Zdravkov}, \citenamefont {Morari}, \citenamefont {Socrovisciuc},
  \citenamefont {Lenk}, \citenamefont {Horn}, \citenamefont {Tagirov},
  \citenamefont {Zaikin}, \citenamefont {Sidorenko}, \citenamefont {Hahn},\
  and\ \citenamefont {Tidecks}}]{Antropov2013}%
  \BibitemOpen
  \bibfield  {author} {\bibinfo {author} {\bibfnamefont {E.}~\bibnamefont
  {Antropov}}, \bibinfo {author} {\bibfnamefont {M.~S.}\ \bibnamefont
  {Kalenkov}}, \bibinfo {author} {\bibfnamefont {J.}~\bibnamefont {Kehrle}},
  \bibinfo {author} {\bibfnamefont {V.~I.}\ \bibnamefont {Zdravkov}}, \bibinfo
  {author} {\bibfnamefont {R.}~\bibnamefont {Morari}}, \bibinfo {author}
  {\bibfnamefont {A.}~\bibnamefont {Socrovisciuc}}, \bibinfo {author}
  {\bibfnamefont {D.}~\bibnamefont {Lenk}}, \bibinfo {author} {\bibfnamefont
  {S.}~\bibnamefont {Horn}}, \bibinfo {author} {\bibfnamefont {L.~R.}\
  \bibnamefont {Tagirov}}, \bibinfo {author} {\bibfnamefont {A.~D.}\
  \bibnamefont {Zaikin}}, \bibinfo {author} {\bibfnamefont {A.~S.}\
  \bibnamefont {Sidorenko}}, \bibinfo {author} {\bibfnamefont {H.}~\bibnamefont
  {Hahn}}, \ and\ \bibinfo {author} {\bibfnamefont {R.}~\bibnamefont
  {Tidecks}},\ }\href {\doibase 10.1088/0953-2048/26/8/085003} {\bibfield
  {journal} {\bibinfo  {journal} {Supercond. Sci. Technol.}\ }\textbf {\bibinfo
  {volume} {26}},\ \bibinfo {pages} {085003} (\bibinfo {year}
  {2013})}\BibitemShut {NoStop}%
\bibitem [{\citenamefont {Du}\ \emph {et~al.}(1995)\citenamefont {Du},
  \citenamefont {Gunzburger},\ and\ \citenamefont {Peterson}}]{Du1995}%
  \BibitemOpen
  \bibfield  {author} {\bibinfo {author} {\bibfnamefont {Q.}~\bibnamefont
  {Du}}, \bibinfo {author} {\bibfnamefont {M.~D.}\ \bibnamefont {Gunzburger}},
  \ and\ \bibinfo {author} {\bibfnamefont {J.~S.}\ \bibnamefont {Peterson}},\
  }\href {\doibase 10.1103/PhysRevB.51.16194} {\bibfield  {journal} {\bibinfo
  {journal} {Phys. Rev. B}\ }\textbf {\bibinfo {volume} {51}},\ \bibinfo
  {pages} {16194} (\bibinfo {year} {1995})}\BibitemShut {NoStop}%
\bibitem [{\citenamefont {Chapman}\ \emph {et~al.}(1996)\citenamefont
  {Chapman}, \citenamefont {Du},\ and\ \citenamefont
  {Gunzburger}}]{Chapman1996}%
  \BibitemOpen
  \bibfield  {author} {\bibinfo {author} {\bibfnamefont {S.~J.}\ \bibnamefont
  {Chapman}}, \bibinfo {author} {\bibfnamefont {Q.}~\bibnamefont {Du}}, \ and\
  \bibinfo {author} {\bibfnamefont {M.~D.}\ \bibnamefont {Gunzburger}},\ }\href
  {\doibase 10.1007/BF00916647} {\bibfield  {journal} {\bibinfo  {journal} {Z.
  Angew. Math. Phys.}\ }\textbf {\bibinfo {volume} {47}},\ \bibinfo {pages}
  {410} (\bibinfo {year} {1996})}\BibitemShut {NoStop}%
\bibitem [{\citenamefont {Sardella}\ and\ \citenamefont
  {Brandt}(2009)}]{Sardella2009}%
  \BibitemOpen
  \bibfield  {author} {\bibinfo {author} {\bibfnamefont {E.}~\bibnamefont
  {Sardella}}\ and\ \bibinfo {author} {\bibfnamefont {E.~H.}\ \bibnamefont
  {Brandt}},\ }\href {\doibase 10.1088/0953-2048/23/2/025015} {\bibfield
  {journal} {\bibinfo  {journal} {Supercond. Sci. Technol.}\ }\textbf {\bibinfo
  {volume} {23}},\ \bibinfo {pages} {025015} (\bibinfo {year}
  {2009})}\BibitemShut {NoStop}%
\bibitem [{\citenamefont {Lu}\ \emph {et~al.}(2016)\citenamefont {Lu},
  \citenamefont {Jing}, \citenamefont {Yong},\ and\ \citenamefont
  {Zhou}}]{LuZhou2016}%
  \BibitemOpen
  \bibfield  {author} {\bibinfo {author} {\bibfnamefont {Y.}~\bibnamefont
  {Lu}}, \bibinfo {author} {\bibfnamefont {Z.}~\bibnamefont {Jing}}, \bibinfo
  {author} {\bibfnamefont {H.}~\bibnamefont {Yong}}, \ and\ \bibinfo {author}
  {\bibfnamefont {Y.}~\bibnamefont {Zhou}},\ }\href {\doibase
  10.1098/rspa.2016.0469} {\bibfield  {journal} {\bibinfo  {journal} {Proc. R.
  Soc. A}\ }\textbf {\bibinfo {volume} {472}},\ \bibinfo {pages} {20160469}
  (\bibinfo {year} {2016})}\BibitemShut {NoStop}%
\bibitem [{\citenamefont {Sabatino}\ \emph {et~al.}(2012)\citenamefont
  {Sabatino}, \citenamefont {Carapella},\ and\ \citenamefont
  {Gombos}}]{Sabatino2012}%
  \BibitemOpen
  \bibfield  {author} {\bibinfo {author} {\bibfnamefont {P.}~\bibnamefont
  {Sabatino}}, \bibinfo {author} {\bibfnamefont {G.}~\bibnamefont {Carapella}},
  \ and\ \bibinfo {author} {\bibfnamefont {M.}~\bibnamefont {Gombos}},\ }\href
  {\doibase 10.1063/1.4759206} {\bibfield  {journal} {\bibinfo  {journal} {J.
  Appl. Phys.}\ }\textbf {\bibinfo {volume} {112}},\ \bibinfo {pages} {083909}
  (\bibinfo {year} {2012})}\BibitemShut {NoStop}%
\bibitem [{\citenamefont {Gladilin}\ \emph {et~al.}(2015)\citenamefont
  {Gladilin}, \citenamefont {Ge}, \citenamefont {Gutierrez}, \citenamefont
  {Timmermans}, \citenamefont {Van~de Vondel}, \citenamefont {Tempere},
  \citenamefont {Devreese},\ and\ \citenamefont {Moshchalkov}}]{Gladilin2015}%
  \BibitemOpen
  \bibfield  {author} {\bibinfo {author} {\bibfnamefont {V.~N.}\ \bibnamefont
  {Gladilin}}, \bibinfo {author} {\bibfnamefont {J.}~\bibnamefont {Ge}},
  \bibinfo {author} {\bibfnamefont {J.}~\bibnamefont {Gutierrez}}, \bibinfo
  {author} {\bibfnamefont {M.}~\bibnamefont {Timmermans}}, \bibinfo {author}
  {\bibfnamefont {J.}~\bibnamefont {Van~de Vondel}}, \bibinfo {author}
  {\bibfnamefont {J.}~\bibnamefont {Tempere}}, \bibinfo {author} {\bibfnamefont
  {J.~T.}\ \bibnamefont {Devreese}}, \ and\ \bibinfo {author} {\bibfnamefont
  {V.~V.}\ \bibnamefont {Moshchalkov}},\ }\href@noop {} {\bibfield  {journal}
  {\bibinfo  {journal} {New J. Phys.}\ }\textbf {\bibinfo {volume} {17}}
  (\bibinfo {year} {2015})}\BibitemShut {NoStop}%
\bibitem [{\citenamefont {Hengstberger}\ \emph {et~al.}(2010)\citenamefont
  {Hengstberger}, \citenamefont {Eisterer},\ and\ \citenamefont
  {Weber}}]{Hengstberger2010}%
  \BibitemOpen
  \bibfield  {author} {\bibinfo {author} {\bibfnamefont {F.}~\bibnamefont
  {Hengstberger}}, \bibinfo {author} {\bibfnamefont {M.}~\bibnamefont
  {Eisterer}}, \ and\ \bibinfo {author} {\bibfnamefont {H.~W.}\ \bibnamefont
  {Weber}},\ }\href@noop {} {\bibfield  {journal} {\bibinfo  {journal} {Appl.
  Phys. Lett.}\ }\textbf {\bibinfo {volume} {96}} (\bibinfo {year} {2010})},\
  \Eprint {http://arxiv.org/abs/1001.1056} {1001.1056} \BibitemShut {NoStop}%
\bibitem [{\citenamefont {Mogro-Campero}\ \emph {et~al.}(1990)\citenamefont
  {Mogro-Campero}, \citenamefont {Turner}, \citenamefont {Hall}, \citenamefont
  {Lewis}, \citenamefont {Peluso},\ and\ \citenamefont
  {Balz}}]{Mogro-Campero1990}%
  \BibitemOpen
  \bibfield  {author} {\bibinfo {author} {\bibfnamefont {A.}~\bibnamefont
  {Mogro-Campero}}, \bibinfo {author} {\bibfnamefont {L.~G.}\ \bibnamefont
  {Turner}}, \bibinfo {author} {\bibfnamefont {E.~L.}\ \bibnamefont {Hall}},
  \bibinfo {author} {\bibfnamefont {N.}~\bibnamefont {Lewis}}, \bibinfo
  {author} {\bibfnamefont {L.~A.}\ \bibnamefont {Peluso}}, \ and\ \bibinfo
  {author} {\bibfnamefont {W.~E.}\ \bibnamefont {Balz}},\ }\href {\doibase
  10.1088/0953-2048/3/2/002} {\bibfield  {journal} {\bibinfo  {journal}
  {Supercond. Sci. Technol.}\ }\textbf {\bibinfo {volume} {3}},\ \bibinfo
  {pages} {62} (\bibinfo {year} {1990})}\BibitemShut {NoStop}%
\bibitem [{\citenamefont {Foltyn}\ \emph {et~al.}(2003)\citenamefont {Foltyn},
  \citenamefont {Arendt}, \citenamefont {Jia}, \citenamefont {Wang},
  \citenamefont {MacManus-Driscoll}, \citenamefont {Kreiskott}, \citenamefont
  {DePaula}, \citenamefont {Stan}, \citenamefont {Groves},\ and\ \citenamefont
  {Dowden}}]{Foltyn2003}%
  \BibitemOpen
  \bibfield  {author} {\bibinfo {author} {\bibfnamefont {S.~R.}\ \bibnamefont
  {Foltyn}}, \bibinfo {author} {\bibfnamefont {P.~N.}\ \bibnamefont {Arendt}},
  \bibinfo {author} {\bibfnamefont {Q.~X.}\ \bibnamefont {Jia}}, \bibinfo
  {author} {\bibfnamefont {H.}~\bibnamefont {Wang}}, \bibinfo {author}
  {\bibfnamefont {J.~L.}\ \bibnamefont {MacManus-Driscoll}}, \bibinfo {author}
  {\bibfnamefont {S.}~\bibnamefont {Kreiskott}}, \bibinfo {author}
  {\bibfnamefont {R.~F.}\ \bibnamefont {DePaula}}, \bibinfo {author}
  {\bibfnamefont {L.}~\bibnamefont {Stan}}, \bibinfo {author} {\bibfnamefont
  {J.~R.}\ \bibnamefont {Groves}}, \ and\ \bibinfo {author} {\bibfnamefont
  {P.~C.}\ \bibnamefont {Dowden}},\ }\href {\doibase 10.1063/1.1584783}
  {\bibfield  {journal} {\bibinfo  {journal} {Appl. Phys. Lett.}\ }\textbf
  {\bibinfo {volume} {82}},\ \bibinfo {pages} {4519} (\bibinfo {year}
  {2003})}\BibitemShut {NoStop}%
\bibitem [{\citenamefont {Onori}\ and\ \citenamefont
  {Rogani}(1985)}]{Onori1985}%
  \BibitemOpen
  \bibfield  {author} {\bibinfo {author} {\bibfnamefont {S.}~\bibnamefont
  {Onori}}\ and\ \bibinfo {author} {\bibfnamefont {A.}~\bibnamefont {Rogani}},\
  }\href {\doibase https://doi.org/10.1016/0378-4363(85)90067-1} {\bibfield
  {journal} {\bibinfo  {journal} {Physica B+C}\ }\textbf {\bibinfo {volume}
  {132}},\ \bibinfo {pages} {217} (\bibinfo {year} {1985})}\BibitemShut
  {NoStop}%
\bibitem [{\citenamefont {Chaudhari}\ and\ \citenamefont
  {Brown}(1965)}]{Chaudhari1965}%
  \BibitemOpen
  \bibfield  {author} {\bibinfo {author} {\bibfnamefont {R.~D.}\ \bibnamefont
  {Chaudhari}}\ and\ \bibinfo {author} {\bibfnamefont {J.~B.}\ \bibnamefont
  {Brown}},\ }\href {\doibase 10.1103/PhysRev.139.A1482} {\bibfield  {journal}
  {\bibinfo  {journal} {Phys. Rev.}\ }\textbf {\bibinfo {volume} {139}},\
  \bibinfo {pages} {A1482} (\bibinfo {year} {1965})}\BibitemShut {NoStop}%
\bibitem [{\citenamefont {Il’in}\ \emph {et~al.}(2010)\citenamefont
  {Il’in}, \citenamefont {Rall}, \citenamefont {Siegel}, \citenamefont
  {Engel}, \citenamefont {Schilling}, \citenamefont {Semenov},\ and\
  \citenamefont {Huebers}}]{Ilin2010}%
  \BibitemOpen
  \bibfield  {author} {\bibinfo {author} {\bibfnamefont {K.}~\bibnamefont
  {Il’in}}, \bibinfo {author} {\bibfnamefont {D.}~\bibnamefont {Rall}},
  \bibinfo {author} {\bibfnamefont {M.}~\bibnamefont {Siegel}}, \bibinfo
  {author} {\bibfnamefont {A.}~\bibnamefont {Engel}}, \bibinfo {author}
  {\bibfnamefont {A.}~\bibnamefont {Schilling}}, \bibinfo {author}
  {\bibfnamefont {A.}~\bibnamefont {Semenov}}, \ and\ \bibinfo {author}
  {\bibfnamefont {H.-W.}\ \bibnamefont {Huebers}},\ }\href {\doibase
  https://doi.org/10.1016/j.physc.2010.02.042} {\bibfield  {journal} {\bibinfo
  {journal} {Physica C}\ }\textbf {\bibinfo {volume} {470}},\ \bibinfo {pages}
  {953} (\bibinfo {year} {2010})}\BibitemShut {NoStop}%
\bibitem [{\citenamefont {Talantsev}\ and\ \citenamefont
  {Tallon}(2015)}]{Talantsev2015}%
  \BibitemOpen
  \bibfield  {author} {\bibinfo {author} {\bibfnamefont {E.~F.}\ \bibnamefont
  {Talantsev}}\ and\ \bibinfo {author} {\bibfnamefont {J.~L.}\ \bibnamefont
  {Tallon}},\ }\href {\doibase 10.1038/ncomms8820} {\bibfield  {journal}
  {\bibinfo  {journal} {Nat. Commun.}\ }\textbf {\bibinfo {volume} {6}},\
  \bibinfo {pages} {7820} (\bibinfo {year} {2015})}\BibitemShut {NoStop}%
\bibitem [{\citenamefont {Brisbois}\ \emph {et~al.}(2017)\citenamefont
  {Brisbois}, \citenamefont {Gladilin}, \citenamefont {Tempere}, \citenamefont
  {Devreese}, \citenamefont {Moshchalkov}, \citenamefont {Colauto},
  \citenamefont {Motta}, \citenamefont {Johansen}, \citenamefont {Fritzsche},
  \citenamefont {Adami}, \citenamefont {Nguyen}, \citenamefont {Ortiz},
  \citenamefont {Kramer},\ and\ \citenamefont {Silhanek}}]{Brisbois2017}%
  \BibitemOpen
  \bibfield  {author} {\bibinfo {author} {\bibfnamefont {J.}~\bibnamefont
  {Brisbois}}, \bibinfo {author} {\bibfnamefont {V.~N.}\ \bibnamefont
  {Gladilin}}, \bibinfo {author} {\bibfnamefont {J.}~\bibnamefont {Tempere}},
  \bibinfo {author} {\bibfnamefont {J.~T.}\ \bibnamefont {Devreese}}, \bibinfo
  {author} {\bibfnamefont {V.~V.}\ \bibnamefont {Moshchalkov}}, \bibinfo
  {author} {\bibfnamefont {F.}~\bibnamefont {Colauto}}, \bibinfo {author}
  {\bibfnamefont {M.}~\bibnamefont {Motta}}, \bibinfo {author} {\bibfnamefont
  {T.~H.}\ \bibnamefont {Johansen}}, \bibinfo {author} {\bibfnamefont
  {J.}~\bibnamefont {Fritzsche}}, \bibinfo {author} {\bibfnamefont {O.-A.}\
  \bibnamefont {Adami}}, \bibinfo {author} {\bibfnamefont {N.~D.}\ \bibnamefont
  {Nguyen}}, \bibinfo {author} {\bibfnamefont {W.~A.}\ \bibnamefont {Ortiz}},
  \bibinfo {author} {\bibfnamefont {R.~B.~G.}\ \bibnamefont {Kramer}}, \ and\
  \bibinfo {author} {\bibfnamefont {A.~V.}\ \bibnamefont {Silhanek}},\ }\href
  {\doibase 10.1103/PhysRevB.95.094506} {\bibfield  {journal} {\bibinfo
  {journal} {Phys. Rev. B}\ }\textbf {\bibinfo {volume} {95}},\ \bibinfo
  {pages} {094506} (\bibinfo {year} {2017})}\BibitemShut {NoStop}%
\bibitem [{\citenamefont {Bean}(1962)}]{bean1962prl}%
  \BibitemOpen
  \bibfield  {author} {\bibinfo {author} {\bibfnamefont {C.~P.}\ \bibnamefont
  {Bean}},\ }\href@noop {} {\bibfield  {journal} {\bibinfo  {journal} {Phys.
  Rev. Lett.}\ }\textbf {\bibinfo {volume} {8}},\ \bibinfo {pages} {250}
  (\bibinfo {year} {1962})}\BibitemShut {NoStop}%
\bibitem [{\citenamefont {Kim}\ \emph {et~al.}(1963)\citenamefont {Kim},
  \citenamefont {Hempstead},\ and\ \citenamefont {Strnad}}]{Kim1963}%
  \BibitemOpen
  \bibfield  {author} {\bibinfo {author} {\bibfnamefont {Y.~B.}\ \bibnamefont
  {Kim}}, \bibinfo {author} {\bibfnamefont {C.~F.}\ \bibnamefont {Hempstead}},
  \ and\ \bibinfo {author} {\bibfnamefont {A.~R.}\ \bibnamefont {Strnad}},\
  }\href {\doibase 10.1103/PhysRev.129.528} {\bibfield  {journal} {\bibinfo
  {journal} {Phys. Rev.}\ }\textbf {\bibinfo {volume} {129}},\ \bibinfo {pages}
  {528} (\bibinfo {year} {1963})}\BibitemShut {NoStop}%
\bibitem [{\citenamefont {Fietz}\ \emph {et~al.}(1964)\citenamefont {Fietz},
  \citenamefont {Beasley}, \citenamefont {Silcox},\ and\ \citenamefont
  {Webb}}]{Fietz1964}%
  \BibitemOpen
  \bibfield  {author} {\bibinfo {author} {\bibfnamefont {W.~A.}\ \bibnamefont
  {Fietz}}, \bibinfo {author} {\bibfnamefont {M.~R.}\ \bibnamefont {Beasley}},
  \bibinfo {author} {\bibfnamefont {J.}~\bibnamefont {Silcox}}, \ and\ \bibinfo
  {author} {\bibfnamefont {W.~W.}\ \bibnamefont {Webb}},\ }\href {\doibase
  10.1103/PhysRev.136.A335} {\bibfield  {journal} {\bibinfo  {journal} {Phys.
  Rev.}\ }\textbf {\bibinfo {volume} {136}},\ \bibinfo {pages} {A335} (\bibinfo
  {year} {1964})}\BibitemShut {NoStop}%
\bibitem [{\citenamefont {Burger}\ \emph {et~al.}(2019)\citenamefont {Burger},
  \citenamefont {Veshchunov}, \citenamefont {Tamegai}, \citenamefont
  {Silhanek}, \citenamefont {Nagasawa}, \citenamefont {Hidaka},\ and\
  \citenamefont {Vanderheyden}}]{burger2019numerical}%
  \BibitemOpen
  \bibfield  {author} {\bibinfo {author} {\bibfnamefont {L.}~\bibnamefont
  {Burger}}, \bibinfo {author} {\bibfnamefont {I.~S.}\ \bibnamefont
  {Veshchunov}}, \bibinfo {author} {\bibfnamefont {T.}~\bibnamefont {Tamegai}},
  \bibinfo {author} {\bibfnamefont {A.}~\bibnamefont {Silhanek}}, \bibinfo
  {author} {\bibfnamefont {S.}~\bibnamefont {Nagasawa}}, \bibinfo {author}
  {\bibfnamefont {M.}~\bibnamefont {Hidaka}}, \ and\ \bibinfo {author}
  {\bibfnamefont {B.}~\bibnamefont {Vanderheyden}},\ }\href@noop {} {\bibfield
  {journal} {\bibinfo  {journal} {Supercond. Sci. Technol.}\ }\textbf {\bibinfo
  {volume} {32}},\ \bibinfo {pages} {125010} (\bibinfo {year}
  {2019})}\BibitemShut {NoStop}%
\bibitem [{\citenamefont {Jiang}\ \emph {et~al.}(2020)\citenamefont {Jiang},
  \citenamefont {Xue}, \citenamefont {Burger}, \citenamefont {Vanderheyden},
  \citenamefont {Silhanek},\ and\ \citenamefont {Zhou}}]{jiang2020selective}%
  \BibitemOpen
  \bibfield  {author} {\bibinfo {author} {\bibfnamefont {L.}~\bibnamefont
  {Jiang}}, \bibinfo {author} {\bibfnamefont {C.}~\bibnamefont {Xue}}, \bibinfo
  {author} {\bibfnamefont {L.}~\bibnamefont {Burger}}, \bibinfo {author}
  {\bibfnamefont {B.}~\bibnamefont {Vanderheyden}}, \bibinfo {author}
  {\bibfnamefont {A.~V.}\ \bibnamefont {Silhanek}}, \ and\ \bibinfo {author}
  {\bibfnamefont {Y.-H.}\ \bibnamefont {Zhou}},\ }\href@noop {} {\bibfield
  {journal} {\bibinfo  {journal} {Phys. Rev. B}\ }\textbf {\bibinfo {volume}
  {101}},\ \bibinfo {pages} {224505} (\bibinfo {year} {2020})}\BibitemShut
  {NoStop}%
\bibitem [{\citenamefont {Motta}\ \emph {et~al.}(2021)\citenamefont {Motta},
  \citenamefont {Burger}, \citenamefont {Jiang}, \citenamefont {Acosta},
  \citenamefont {Jeli{\'c}}, \citenamefont {Colauto}, \citenamefont {Ortiz},
  \citenamefont {Johansen}, \citenamefont {Milo{\v{s}}evi{\'c}}, \citenamefont
  {Cirillo}, \citenamefont {Attanasio}, \citenamefont {Xue}, \citenamefont
  {Silhanek},\ and\ \citenamefont {Vanderheyden}}]{motta2021metamorphosis}%
  \BibitemOpen
  \bibfield  {author} {\bibinfo {author} {\bibfnamefont {M.}~\bibnamefont
  {Motta}}, \bibinfo {author} {\bibfnamefont {L.}~\bibnamefont {Burger}},
  \bibinfo {author} {\bibfnamefont {L.}~\bibnamefont {Jiang}}, \bibinfo
  {author} {\bibfnamefont {J.~D.~G.}\ \bibnamefont {Acosta}}, \bibinfo {author}
  {\bibfnamefont {{\v{Z}}.}~\bibnamefont {Jeli{\'c}}}, \bibinfo {author}
  {\bibfnamefont {F.}~\bibnamefont {Colauto}}, \bibinfo {author} {\bibfnamefont
  {W.~A.}\ \bibnamefont {Ortiz}}, \bibinfo {author} {\bibfnamefont {T.~H.}\
  \bibnamefont {Johansen}}, \bibinfo {author} {\bibfnamefont {M.~V.}\
  \bibnamefont {Milo{\v{s}}evi{\'c}}}, \bibinfo {author} {\bibfnamefont
  {C.}~\bibnamefont {Cirillo}}, \bibinfo {author} {\bibfnamefont
  {C.}~\bibnamefont {Attanasio}}, \bibinfo {author} {\bibfnamefont
  {C.}~\bibnamefont {Xue}}, \bibinfo {author} {\bibfnamefont {A.~V.}\
  \bibnamefont {Silhanek}}, \ and\ \bibinfo {author} {\bibfnamefont
  {B.}~\bibnamefont {Vanderheyden}},\ }\href@noop {} {\bibfield  {journal}
  {\bibinfo  {journal} {Phys. Rev. B}\ }\textbf {\bibinfo {volume} {103}},\
  \bibinfo {pages} {224514} (\bibinfo {year} {2021})}\BibitemShut {NoStop}%
\bibitem [{\citenamefont {Chaves}\ \emph {et~al.}(2021)\citenamefont {Chaves},
  \citenamefont {de~Araújo}, \citenamefont {Carmo}, \citenamefont {Colauto},
  \citenamefont {de~Oliveira}, \citenamefont {de~Andrade}, \citenamefont
  {Johansen}, \citenamefont {Silhanek}, \citenamefont {Ortiz},\ and\
  \citenamefont {Motta}}]{davi2021}%
  \BibitemOpen
  \bibfield  {author} {\bibinfo {author} {\bibfnamefont {D.~A.~D.}\
  \bibnamefont {Chaves}}, \bibinfo {author} {\bibfnamefont {I.~M.}\
  \bibnamefont {de~Araújo}}, \bibinfo {author} {\bibfnamefont
  {D.}~\bibnamefont {Carmo}}, \bibinfo {author} {\bibfnamefont
  {F.}~\bibnamefont {Colauto}}, \bibinfo {author} {\bibfnamefont {A.~A.~M.}\
  \bibnamefont {de~Oliveira}}, \bibinfo {author} {\bibfnamefont {A.~M.~H.}\
  \bibnamefont {de~Andrade}}, \bibinfo {author} {\bibfnamefont {T.~H.}\
  \bibnamefont {Johansen}}, \bibinfo {author} {\bibfnamefont {A.~V.}\
  \bibnamefont {Silhanek}}, \bibinfo {author} {\bibfnamefont {W.~A.}\
  \bibnamefont {Ortiz}}, \ and\ \bibinfo {author} {\bibfnamefont
  {M.}~\bibnamefont {Motta}},\ }\href {\doibase 10.1063/5.0058680} {\bibfield
  {journal} {\bibinfo  {journal} {Appl. Phys. Lett.}\ }\textbf {\bibinfo
  {volume} {119}},\ \bibinfo {pages} {022602} (\bibinfo {year}
  {2021})}\BibitemShut {NoStop}%
\bibitem [{\citenamefont {Brandt}(1997)}]{Brandt1997}%
  \BibitemOpen
  \bibfield  {author} {\bibinfo {author} {\bibfnamefont {E.~H.}\ \bibnamefont
  {Brandt}},\ }\href {\doibase 10.1103/PhysRevB.55.14513} {\bibfield  {journal}
  {\bibinfo  {journal} {Phys. Rev. B}\ }\textbf {\bibinfo {volume} {55}},\
  \bibinfo {pages} {14513} (\bibinfo {year} {1997})}\BibitemShut {NoStop}%
\bibitem [{\citenamefont {Brandt}(1994{\natexlab{a}})}]{Brandt1994}%
  \BibitemOpen
  \bibfield  {author} {\bibinfo {author} {\bibfnamefont {E.~H.}\ \bibnamefont
  {Brandt}},\ }\href {\doibase 10.1103/PhysRevB.49.9024} {\bibfield  {journal}
  {\bibinfo  {journal} {Phys. Rev. B}\ }\textbf {\bibinfo {volume} {49}},\
  \bibinfo {pages} {9024} (\bibinfo {year} {1994}{\natexlab{a}})}\BibitemShut
  {NoStop}%
\bibitem [{\citenamefont {Clem}\ and\ \citenamefont
  {Sanchez}(1994)}]{Clem1994}%
  \BibitemOpen
  \bibfield  {author} {\bibinfo {author} {\bibfnamefont {J.~R.}\ \bibnamefont
  {Clem}}\ and\ \bibinfo {author} {\bibfnamefont {A.}~\bibnamefont {Sanchez}},\
  }\href {\doibase 10.1103/PhysRevB.50.9355} {\bibfield  {journal} {\bibinfo
  {journal} {Phys. Rev. B}\ }\textbf {\bibinfo {volume} {50}},\ \bibinfo
  {pages} {9355} (\bibinfo {year} {1994})}\BibitemShut {NoStop}%
\bibitem [{\citenamefont {Brandt}(1994{\natexlab{b}})}]{Brandt1994a}%
  \BibitemOpen
  \bibfield  {author} {\bibinfo {author} {\bibfnamefont {E.~H.}\ \bibnamefont
  {Brandt}},\ }\href {\doibase 10.1103/PhysRevB.50.4034} {\bibfield  {journal}
  {\bibinfo  {journal} {Phys. Rev. B}\ }\textbf {\bibinfo {volume} {50}},\
  \bibinfo {pages} {4034} (\bibinfo {year} {1994}{\natexlab{b}})}\BibitemShut
  {NoStop}%
\bibitem [{\citenamefont {Zeldov}\ \emph {et~al.}(1994)\citenamefont {Zeldov},
  \citenamefont {Clem}, \citenamefont {McElfresh},\ and\ \citenamefont
  {Darwin}}]{zeldov1994profile}%
  \BibitemOpen
  \bibfield  {author} {\bibinfo {author} {\bibfnamefont {E.}~\bibnamefont
  {Zeldov}}, \bibinfo {author} {\bibfnamefont {J.~R.}\ \bibnamefont {Clem}},
  \bibinfo {author} {\bibfnamefont {M.}~\bibnamefont {McElfresh}}, \ and\
  \bibinfo {author} {\bibfnamefont {M.}~\bibnamefont {Darwin}},\ }\href@noop {}
  {\bibfield  {journal} {\bibinfo  {journal} {Phys. Rev. B}\ }\textbf {\bibinfo
  {volume} {49}},\ \bibinfo {pages} {9802} (\bibinfo {year}
  {1994})}\BibitemShut {NoStop}%
\bibitem [{\citenamefont {Altshuler}\ and\ \citenamefont
  {Johansen}(2004)}]{altshuler_colloquium:_2004}%
  \BibitemOpen
  \bibfield  {author} {\bibinfo {author} {\bibfnamefont {E.}~\bibnamefont
  {Altshuler}}\ and\ \bibinfo {author} {\bibfnamefont {T.~H.}\ \bibnamefont
  {Johansen}},\ }\href {\doibase 10.1103/RevModPhys.76.471} {\bibfield
  {journal} {\bibinfo  {journal} {Rev. Mod.Phys.}\ }\textbf {\bibinfo {volume}
  {76}},\ \bibinfo {pages} {471} (\bibinfo {year} {2004})}\BibitemShut
  {NoStop}%
\bibitem [{\citenamefont {Denisov}\ \emph
  {et~al.}(2006{\natexlab{a}})\citenamefont {Denisov}, \citenamefont
  {Shantsev}, \citenamefont {Galperin}, \citenamefont {Choi}, \citenamefont
  {Lee}, \citenamefont {Lee}, \citenamefont {Bobyl}, \citenamefont {Goa},
  \citenamefont {Olsen},\ and\ \citenamefont {Johansen}}]{Denisov2006}%
  \BibitemOpen
  \bibfield  {author} {\bibinfo {author} {\bibfnamefont {D.~V.}\ \bibnamefont
  {Denisov}}, \bibinfo {author} {\bibfnamefont {D.~V.}\ \bibnamefont
  {Shantsev}}, \bibinfo {author} {\bibfnamefont {Y.~M.}\ \bibnamefont
  {Galperin}}, \bibinfo {author} {\bibfnamefont {E.~M.}\ \bibnamefont {Choi}},
  \bibinfo {author} {\bibfnamefont {H.~S.}\ \bibnamefont {Lee}}, \bibinfo
  {author} {\bibfnamefont {S.~I.}\ \bibnamefont {Lee}}, \bibinfo {author}
  {\bibfnamefont {A.~V.}\ \bibnamefont {Bobyl}}, \bibinfo {author}
  {\bibfnamefont {P.~E.}\ \bibnamefont {Goa}}, \bibinfo {author} {\bibfnamefont
  {A.~A.}\ \bibnamefont {Olsen}}, \ and\ \bibinfo {author} {\bibfnamefont
  {T.~H.}\ \bibnamefont {Johansen}},\ }\href@noop {} {\bibfield  {journal}
  {\bibinfo  {journal} {Phys. Rev. Lett.}\ }\textbf {\bibinfo {volume} {97}}
  (\bibinfo {year} {2006}{\natexlab{a}})}\BibitemShut {NoStop}%
\bibitem [{\citenamefont {Colauto}\ \emph {et~al.}(2020)\citenamefont
  {Colauto}, \citenamefont {Motta},\ and\ \citenamefont
  {Ortiz}}]{colauto2020controlling}%
  \BibitemOpen
  \bibfield  {author} {\bibinfo {author} {\bibfnamefont {F.}~\bibnamefont
  {Colauto}}, \bibinfo {author} {\bibfnamefont {M.}~\bibnamefont {Motta}}, \
  and\ \bibinfo {author} {\bibfnamefont {W.~A.}\ \bibnamefont {Ortiz}},\
  }\href@noop {} {\bibfield  {journal} {\bibinfo  {journal} {Supercond. Sci.
  Technol.}\ }\textbf {\bibinfo {volume} {34}},\ \bibinfo {pages} {013002}
  (\bibinfo {year} {2020})}\BibitemShut {NoStop}%
\bibitem [{\citenamefont {Yurchenko}\ \emph {et~al.}(2009)\citenamefont
  {Yurchenko}, \citenamefont {Johansen},\ and\ \citenamefont
  {Galperin}}]{Yurchenko2009}%
  \BibitemOpen
  \bibfield  {author} {\bibinfo {author} {\bibfnamefont {V.~V.}\ \bibnamefont
  {Yurchenko}}, \bibinfo {author} {\bibfnamefont {T.~H.}\ \bibnamefont
  {Johansen}}, \ and\ \bibinfo {author} {\bibfnamefont {Y.~M.}\ \bibnamefont
  {Galperin}},\ }\href {\doibase 10.1063/1.3224713} {\bibfield  {journal}
  {\bibinfo  {journal} {Low Temp. Phys.}\ }\textbf {\bibinfo {volume} {35}},\
  \bibinfo {pages} {619} (\bibinfo {year} {2009})}\BibitemShut {NoStop}%
\bibitem [{\citenamefont {Vestg{\aa}rden}\ \emph {et~al.}(2018)\citenamefont
  {Vestg{\aa}rden}, \citenamefont {Johansen},\ and\ \citenamefont
  {Galperin}}]{vestgaarden2018nucleation}%
  \BibitemOpen
  \bibfield  {author} {\bibinfo {author} {\bibfnamefont {J.~I.}\ \bibnamefont
  {Vestg{\aa}rden}}, \bibinfo {author} {\bibfnamefont {T.~H.}\ \bibnamefont
  {Johansen}}, \ and\ \bibinfo {author} {\bibfnamefont {Y.~M.}\ \bibnamefont
  {Galperin}},\ }\href@noop {} {\bibfield  {journal} {\bibinfo  {journal} {Low
  Temp. Phys.}\ }\textbf {\bibinfo {volume} {44}},\ \bibinfo {pages} {460}
  (\bibinfo {year} {2018})}\BibitemShut {NoStop}%
\bibitem [{\citenamefont {Vestg\aa{}rden}\ \emph {et~al.}(2011)\citenamefont
  {Vestg\aa{}rden}, \citenamefont {Shantsev}, \citenamefont {Galperin},\ and\
  \citenamefont {Johansen}}]{Vestgarden2011}%
  \BibitemOpen
  \bibfield  {author} {\bibinfo {author} {\bibfnamefont {J.~I.}\ \bibnamefont
  {Vestg\aa{}rden}}, \bibinfo {author} {\bibfnamefont {D.~V.}\ \bibnamefont
  {Shantsev}}, \bibinfo {author} {\bibfnamefont {Y.~M.}\ \bibnamefont
  {Galperin}}, \ and\ \bibinfo {author} {\bibfnamefont {T.~H.}\ \bibnamefont
  {Johansen}},\ }\href {\doibase 10.1103/PhysRevB.84.054537} {\bibfield
  {journal} {\bibinfo  {journal} {Phys. Rev. B}\ }\textbf {\bibinfo {volume}
  {84}},\ \bibinfo {pages} {054537} (\bibinfo {year} {2011})}\BibitemShut
  {NoStop}%
\bibitem [{\citenamefont {Blanco~Alvarez}\ \emph {et~al.}(2019)\citenamefont
  {Blanco~Alvarez}, \citenamefont {Brisbois}, \citenamefont {Melinte},
  \citenamefont {Kramer},\ and\ \citenamefont
  {Silhanek}}]{blanco2019statistics}%
  \BibitemOpen
  \bibfield  {author} {\bibinfo {author} {\bibfnamefont {S.}~\bibnamefont
  {Blanco~Alvarez}}, \bibinfo {author} {\bibfnamefont {J.}~\bibnamefont
  {Brisbois}}, \bibinfo {author} {\bibfnamefont {S.}~\bibnamefont {Melinte}},
  \bibinfo {author} {\bibfnamefont {R.~B.~G.}\ \bibnamefont {Kramer}}, \ and\
  \bibinfo {author} {\bibfnamefont {A.~V.}\ \bibnamefont {Silhanek}},\
  }\href@noop {} {\bibfield  {journal} {\bibinfo  {journal} {Sci. Rep.}\
  }\textbf {\bibinfo {volume} {9}},\ \bibinfo {pages} {3659} (\bibinfo {year}
  {2019})}\BibitemShut {NoStop}%
\bibitem [{\citenamefont {Menghini}\ \emph {et~al.}(2005)\citenamefont
  {Menghini}, \citenamefont {Wijngaarden}, \citenamefont {Silhanek},
  \citenamefont {Raedts},\ and\ \citenamefont {Moshchalkov}}]{Menghini2005}%
  \BibitemOpen
  \bibfield  {author} {\bibinfo {author} {\bibfnamefont {M.}~\bibnamefont
  {Menghini}}, \bibinfo {author} {\bibfnamefont {R.~J.}\ \bibnamefont
  {Wijngaarden}}, \bibinfo {author} {\bibfnamefont {A.~V.}\ \bibnamefont
  {Silhanek}}, \bibinfo {author} {\bibfnamefont {S.}~\bibnamefont {Raedts}}, \
  and\ \bibinfo {author} {\bibfnamefont {V.~V.}\ \bibnamefont {Moshchalkov}},\
  }\href {\doibase 10.1103/PhysRevB.71.104506} {\bibfield  {journal} {\bibinfo
  {journal} {Phys. Rev. B}\ }\textbf {\bibinfo {volume} {71}},\ \bibinfo
  {pages} {1} (\bibinfo {year} {2005})}\BibitemShut {NoStop}%
\bibitem [{\citenamefont {Motta}\ \emph {et~al.}(2014)\citenamefont {Motta},
  \citenamefont {Colauto}, \citenamefont {Vestg{\aa}rden}, \citenamefont
  {Fritzsche}, \citenamefont {Timmermans}, \citenamefont {Cuppens},
  \citenamefont {Attanasio}, \citenamefont {Cirillo}, \citenamefont
  {Moshchalkov}, \citenamefont {Van~de Vondel}, \citenamefont {Johansen},
  \citenamefont {Ortiz},\ and\ \citenamefont {Silhanek}}]{Motta2014}%
  \BibitemOpen
  \bibfield  {author} {\bibinfo {author} {\bibfnamefont {M.}~\bibnamefont
  {Motta}}, \bibinfo {author} {\bibfnamefont {F.}~\bibnamefont {Colauto}},
  \bibinfo {author} {\bibfnamefont {J.~I.}\ \bibnamefont {Vestg{\aa}rden}},
  \bibinfo {author} {\bibfnamefont {J.}~\bibnamefont {Fritzsche}}, \bibinfo
  {author} {\bibfnamefont {M.}~\bibnamefont {Timmermans}}, \bibinfo {author}
  {\bibfnamefont {J.}~\bibnamefont {Cuppens}}, \bibinfo {author} {\bibfnamefont
  {C.}~\bibnamefont {Attanasio}}, \bibinfo {author} {\bibfnamefont
  {C.}~\bibnamefont {Cirillo}}, \bibinfo {author} {\bibfnamefont {V.~V.}\
  \bibnamefont {Moshchalkov}}, \bibinfo {author} {\bibfnamefont
  {J.}~\bibnamefont {Van~de Vondel}}, \bibinfo {author} {\bibfnamefont {T.~H.}\
  \bibnamefont {Johansen}}, \bibinfo {author} {\bibfnamefont {W.~A.}\
  \bibnamefont {Ortiz}}, \ and\ \bibinfo {author} {\bibfnamefont {A.~V.}\
  \bibnamefont {Silhanek}},\ }\href {\doibase 10.1103/PhysRevB.89.134508}
  {\bibfield  {journal} {\bibinfo  {journal} {Phys. Rev. B}\ }\textbf {\bibinfo
  {volume} {89}},\ \bibinfo {pages} {134508} (\bibinfo {year}
  {2014})}\BibitemShut {NoStop}%
\bibitem [{\citenamefont {Abal’osheva}\ \emph {et~al.}(2010)\citenamefont
  {Abal’osheva}, \citenamefont {Abal’oshev}, \citenamefont {Cieplak},
  \citenamefont {Zhu},\ and\ \citenamefont {Chien}}]{Abalosheva2010}%
  \BibitemOpen
  \bibfield  {author} {\bibinfo {author} {\bibfnamefont {I.}~\bibnamefont
  {Abal’osheva}}, \bibinfo {author} {\bibfnamefont {A.}~\bibnamefont
  {Abal’oshev}}, \bibinfo {author} {\bibfnamefont {M.}~\bibnamefont
  {Cieplak}}, \bibinfo {author} {\bibfnamefont {L.}~\bibnamefont {Zhu}}, \ and\
  \bibinfo {author} {\bibfnamefont {C.-L.}\ \bibnamefont {Chien}},\ }\href
  {\doibase 10.12693/APhysPolA.118.396} {\bibfield  {journal} {\bibinfo
  {journal} {Acta Phys. Pol. A}\ }\textbf {\bibinfo {volume} {118}},\ \bibinfo
  {pages} {396} (\bibinfo {year} {2010})}\BibitemShut {NoStop}%
\bibitem [{\citenamefont {Vestgarden}\ \emph {et~al.}(2013)\citenamefont
  {Vestgarden}, \citenamefont {Galperin},\ and\ \citenamefont
  {Johansen}}]{Vestgarden2013arxiv}%
  \BibitemOpen
  \bibfield  {author} {\bibinfo {author} {\bibfnamefont {J.~I.}\ \bibnamefont
  {Vestgarden}}, \bibinfo {author} {\bibfnamefont {Y.~M.}\ \bibnamefont
  {Galperin}}, \ and\ \bibinfo {author} {\bibfnamefont {T.~H.}\ \bibnamefont
  {Johansen}},\ }\href@noop {} {\enquote {\bibinfo {title} {Dendritic flux
  avalanches in superconducting films of different thickness},}\ } (\bibinfo
  {year} {2013}),\ \Eprint {http://arxiv.org/abs/1309.6463} {arXiv:1309.6463
  [cond-mat.supr-con]} \BibitemShut {NoStop}%
\bibitem [{\citenamefont {Abaloszewa}\ \emph {et~al.}(2022)\citenamefont
  {Abaloszewa}, \citenamefont {Cieplak},\ and\ \citenamefont
  {Abaloszew}}]{Abaloszewa2022}%
  \BibitemOpen
  \bibfield  {author} {\bibinfo {author} {\bibfnamefont {I.}~\bibnamefont
  {Abaloszewa}}, \bibinfo {author} {\bibfnamefont {M.~Z.}\ \bibnamefont
  {Cieplak}}, \ and\ \bibinfo {author} {\bibfnamefont {A.}~\bibnamefont
  {Abaloszew}},\ }\href@noop {} {\enquote {\bibinfo {title} {Thermomagnetic
  instabilities in {Nb} films deposited on glass substrates},}\ } (\bibinfo
  {year} {2022}),\ \Eprint {http://arxiv.org/abs/2207.12811} {arXiv:2207.12811
  [cond-mat.supr-con]} \BibitemShut {NoStop}%
\bibitem [{\citenamefont {Helseth}\ \emph {et~al.}(2001)\citenamefont
  {Helseth}, \citenamefont {Hansen}, \citenamefont {Il’yashenko},
  \citenamefont {Baziljevich},\ and\ \citenamefont {Johansen}}]{Helseth2001}%
  \BibitemOpen
  \bibfield  {author} {\bibinfo {author} {\bibfnamefont {L.~E.}\ \bibnamefont
  {Helseth}}, \bibinfo {author} {\bibfnamefont {R.~W.}\ \bibnamefont {Hansen}},
  \bibinfo {author} {\bibfnamefont {E.~I.}\ \bibnamefont {Il’yashenko}},
  \bibinfo {author} {\bibfnamefont {M.}~\bibnamefont {Baziljevich}}, \ and\
  \bibinfo {author} {\bibfnamefont {T.~H.}\ \bibnamefont {Johansen}},\
  }\href@noop {} {\bibfield  {journal} {\bibinfo  {journal} {Phys. Rev. B}\
  }\textbf {\bibinfo {volume} {64}},\ \bibinfo {pages} {174406} (\bibinfo
  {year} {2001})}\BibitemShut {NoStop}%
\bibitem [{\citenamefont {Helseth}\ \emph {et~al.}(2002)\citenamefont
  {Helseth}, \citenamefont {Solovyev}, \citenamefont {Hansen}, \citenamefont
  {Il'yashenko}, \citenamefont {Baziljevich},\ and\ \citenamefont
  {Johansen}}]{Helseth2002}%
  \BibitemOpen
  \bibfield  {author} {\bibinfo {author} {\bibfnamefont {L.~E.}\ \bibnamefont
  {Helseth}}, \bibinfo {author} {\bibfnamefont {A.~G.}\ \bibnamefont
  {Solovyev}}, \bibinfo {author} {\bibfnamefont {R.~W.}\ \bibnamefont
  {Hansen}}, \bibinfo {author} {\bibfnamefont {E.~I.}\ \bibnamefont
  {Il'yashenko}}, \bibinfo {author} {\bibfnamefont {M.}~\bibnamefont
  {Baziljevich}}, \ and\ \bibinfo {author} {\bibfnamefont {T.~H.}\ \bibnamefont
  {Johansen}},\ }\href {\doibase 10.1103/PhysRevB.66.064405} {\bibfield
  {journal} {\bibinfo  {journal} {Phys. Rev. B}\ }\textbf {\bibinfo {volume}
  {66}},\ \bibinfo {pages} {1} (\bibinfo {year} {2002})}\BibitemShut {NoStop}%
\bibitem [{\citenamefont {Shaw}\ \emph {et~al.}(2018)\citenamefont {Shaw},
  \citenamefont {Brisbois}, \citenamefont {Pinheiro}, \citenamefont
  {M{\"u}ller}, \citenamefont {Blanco~Alvarez}, \citenamefont {Devillers},
  \citenamefont {Dempsey}, \citenamefont {Scheerder}, \citenamefont {Van~de
  Vondel}, \citenamefont {Melinte}, \citenamefont {Vanderbemden}, \citenamefont
  {Motta}, \citenamefont {Ortiz}, \citenamefont {Hasselbach}, \citenamefont
  {Kramer},\ and\ \citenamefont {Silhanek}}]{shaw2018quantitative}%
  \BibitemOpen
  \bibfield  {author} {\bibinfo {author} {\bibfnamefont {G.}~\bibnamefont
  {Shaw}}, \bibinfo {author} {\bibfnamefont {J.}~\bibnamefont {Brisbois}},
  \bibinfo {author} {\bibfnamefont {L.~B. G.~L.}\ \bibnamefont {Pinheiro}},
  \bibinfo {author} {\bibfnamefont {J.}~\bibnamefont {M{\"u}ller}}, \bibinfo
  {author} {\bibfnamefont {S.}~\bibnamefont {Blanco~Alvarez}}, \bibinfo
  {author} {\bibfnamefont {T.}~\bibnamefont {Devillers}}, \bibinfo {author}
  {\bibfnamefont {N.~M.}\ \bibnamefont {Dempsey}}, \bibinfo {author}
  {\bibfnamefont {J.~E.}\ \bibnamefont {Scheerder}}, \bibinfo {author}
  {\bibfnamefont {J.}~\bibnamefont {Van~de Vondel}}, \bibinfo {author}
  {\bibfnamefont {S.}~\bibnamefont {Melinte}}, \bibinfo {author} {\bibfnamefont
  {P.}~\bibnamefont {Vanderbemden}}, \bibinfo {author} {\bibfnamefont
  {M.}~\bibnamefont {Motta}}, \bibinfo {author} {\bibfnamefont {W.~A.}\
  \bibnamefont {Ortiz}}, \bibinfo {author} {\bibfnamefont {K.}~\bibnamefont
  {Hasselbach}}, \bibinfo {author} {\bibfnamefont {R.~B.~G.}\ \bibnamefont
  {Kramer}}, \ and\ \bibinfo {author} {\bibfnamefont {A.~V.}\ \bibnamefont
  {Silhanek}},\ }\href@noop {} {\bibfield  {journal} {\bibinfo  {journal} {Rev.
  Sci. Instrum.}\ }\textbf {\bibinfo {volume} {89}},\ \bibinfo {pages} {023705}
  (\bibinfo {year} {2018})}\BibitemShut {NoStop}%
\bibitem [{\citenamefont {Th{\'{e}}venaz}\ \emph {et~al.}(1998)\citenamefont
  {Th{\'{e}}venaz}, \citenamefont {Ruttimann},\ and\ \citenamefont
  {Unser}}]{stackreg}%
  \BibitemOpen
  \bibfield  {author} {\bibinfo {author} {\bibfnamefont {P.}~\bibnamefont
  {Th{\'{e}}venaz}}, \bibinfo {author} {\bibfnamefont {U.~E.}\ \bibnamefont
  {Ruttimann}}, \ and\ \bibinfo {author} {\bibfnamefont {M.}~\bibnamefont
  {Unser}},\ }\href@noop {} {\bibfield  {journal} {\bibinfo  {journal} {IEEE
  Trans. Image Process.}\ }\textbf {\bibinfo {volume} {7}},\ \bibinfo {pages}
  {27} (\bibinfo {year} {1998})}\BibitemShut {NoStop}%
\bibitem [{\citenamefont {Schneider}\ \emph {et~al.}(2012)\citenamefont
  {Schneider}, \citenamefont {Rasband},\ and\ \citenamefont
  {Eliceiri}}]{schneider2012nih}%
  \BibitemOpen
  \bibfield  {author} {\bibinfo {author} {\bibfnamefont {C.~A.}\ \bibnamefont
  {Schneider}}, \bibinfo {author} {\bibfnamefont {W.~S.}\ \bibnamefont
  {Rasband}}, \ and\ \bibinfo {author} {\bibfnamefont {K.~W.}\ \bibnamefont
  {Eliceiri}},\ }\href@noop {} {\bibfield  {journal} {\bibinfo  {journal} {Nat.
  Methods}\ }\textbf {\bibinfo {volume} {9}},\ \bibinfo {pages} {671} (\bibinfo
  {year} {2012})}\BibitemShut {NoStop}%
\bibitem [{\citenamefont {Meltzer}\ \emph {et~al.}(2017)\citenamefont
  {Meltzer}, \citenamefont {Levin},\ and\ \citenamefont
  {Zeldov}}]{meltzer2017reconstruction}%
  \BibitemOpen
  \bibfield  {author} {\bibinfo {author} {\bibfnamefont {A.~Y.}\ \bibnamefont
  {Meltzer}}, \bibinfo {author} {\bibfnamefont {E.}~\bibnamefont {Levin}}, \
  and\ \bibinfo {author} {\bibfnamefont {E.}~\bibnamefont {Zeldov}},\ }\href
  {\doibase 10.1103/PhysRevApplied.8.064030} {\bibfield  {journal} {\bibinfo
  {journal} {Phys. Rev. Applied}\ }\textbf {\bibinfo {volume} {8}},\ \bibinfo
  {pages} {064030} (\bibinfo {year} {2017})}\BibitemShut {NoStop}%
\bibitem [{\citenamefont {Denisov}\ \emph
  {et~al.}(2006{\natexlab{b}})\citenamefont {Denisov}, \citenamefont
  {Rakhmanov}, \citenamefont {Shantsev}, \citenamefont {Galperin},\ and\
  \citenamefont {Johansen}}]{Denisov2006a}%
  \BibitemOpen
  \bibfield  {author} {\bibinfo {author} {\bibfnamefont {D.~V.}\ \bibnamefont
  {Denisov}}, \bibinfo {author} {\bibfnamefont {A.~L.}\ \bibnamefont
  {Rakhmanov}}, \bibinfo {author} {\bibfnamefont {D.~V.}\ \bibnamefont
  {Shantsev}}, \bibinfo {author} {\bibfnamefont {Y.~M.}\ \bibnamefont
  {Galperin}}, \ and\ \bibinfo {author} {\bibfnamefont {T.~H.}\ \bibnamefont
  {Johansen}},\ }\href {\doibase 10.1103/PhysRevB.73.014512} {\bibfield
  {journal} {\bibinfo  {journal} {Phys. Rev. B}\ }\textbf {\bibinfo {volume}
  {73}},\ \bibinfo {pages} {014512} (\bibinfo {year}
  {2006}{\natexlab{b}})}\BibitemShut {NoStop}%
\bibitem [{\citenamefont {Mints}\ and\ \citenamefont
  {Rakhmanov}(1981)}]{mints_critical_1981}%
  \BibitemOpen
  \bibfield  {author} {\bibinfo {author} {\bibfnamefont {R.}~\bibnamefont
  {Mints}}\ and\ \bibinfo {author} {\bibfnamefont {A.}~\bibnamefont
  {Rakhmanov}},\ }\href {\doibase 10.1103/RevModPhys.53.551} {\bibfield
  {journal} {\bibinfo  {journal} {Rev. Mod. Phys.}\ }\textbf {\bibinfo {volume}
  {53}},\ \bibinfo {pages} {551} (\bibinfo {year} {1981})}\BibitemShut
  {NoStop}%
\bibitem [{\citenamefont {Portela}\ \emph {et~al.}(2015)\citenamefont
  {Portela}, \citenamefont {Corredor}, \citenamefont {Barrozo}, \citenamefont
  {Jung}, \citenamefont {Zhang}, \citenamefont {Vanacken}, \citenamefont
  {Moshchalkov},\ and\ \citenamefont {Aguiar}}]{Portela2015}%
  \BibitemOpen
  \bibfield  {author} {\bibinfo {author} {\bibfnamefont {F.~S.}\ \bibnamefont
  {Portela}}, \bibinfo {author} {\bibfnamefont {L.~T.}\ \bibnamefont
  {Corredor}}, \bibinfo {author} {\bibfnamefont {P.}~\bibnamefont {Barrozo}},
  \bibinfo {author} {\bibfnamefont {S.-G.}\ \bibnamefont {Jung}}, \bibinfo
  {author} {\bibfnamefont {G.}~\bibnamefont {Zhang}}, \bibinfo {author}
  {\bibfnamefont {J.}~\bibnamefont {Vanacken}}, \bibinfo {author}
  {\bibfnamefont {V.~V.}\ \bibnamefont {Moshchalkov}}, \ and\ \bibinfo {author}
  {\bibfnamefont {J.~A.}\ \bibnamefont {Aguiar}},\ }\href {\doibase
  10.1088/0953-2048/28/3/034001} {\bibfield  {journal} {\bibinfo  {journal}
  {Superconductor Science and Technology}\ }\textbf {\bibinfo {volume} {28}},\
  \bibinfo {pages} {034001} (\bibinfo {year} {2015})}\BibitemShut {NoStop}%
\bibitem [{\citenamefont {Bean}(1964)}]{bean1964rmp}%
  \BibitemOpen
  \bibfield  {author} {\bibinfo {author} {\bibfnamefont {C.~P.}\ \bibnamefont
  {Bean}},\ }\href@noop {} {\bibfield  {journal} {\bibinfo  {journal} {Rev.
  Mod. Phys.}\ }\textbf {\bibinfo {volume} {36}},\ \bibinfo {pages} {31}
  (\bibinfo {year} {1964})}\BibitemShut {NoStop}%
\bibitem [{\citenamefont {Schuster}\ \emph {et~al.}(1996)\citenamefont
  {Schuster}, \citenamefont {Kuhn},\ and\ \citenamefont
  {Brandt}}]{Schuster1996}%
  \BibitemOpen
  \bibfield  {author} {\bibinfo {author} {\bibfnamefont {T.}~\bibnamefont
  {Schuster}}, \bibinfo {author} {\bibfnamefont {H.}~\bibnamefont {Kuhn}}, \
  and\ \bibinfo {author} {\bibfnamefont {E.~H.}\ \bibnamefont {Brandt}},\
  }\href {\doibase 10.1103/PhysRevB.54.3514} {\bibfield  {journal} {\bibinfo
  {journal} {Phys. Rev. B}\ }\textbf {\bibinfo {volume} {54}},\ \bibinfo
  {pages} {3514} (\bibinfo {year} {1996})}\BibitemShut {NoStop}%
\bibitem [{\citenamefont {Valerio-Cuadros}\ \emph {et~al.}(2021)\citenamefont
  {Valerio-Cuadros}, \citenamefont {Chaves}, \citenamefont {Colauto},
  \citenamefont {Oliveira}, \citenamefont {Andrade}, \citenamefont {Johansen},
  \citenamefont {Ortiz},\ and\ \citenamefont
  {Motta}}]{valerio2021superconducting}%
  \BibitemOpen
  \bibfield  {author} {\bibinfo {author} {\bibfnamefont {M.~I.}\ \bibnamefont
  {Valerio-Cuadros}}, \bibinfo {author} {\bibfnamefont {D.~A.~D.}\ \bibnamefont
  {Chaves}}, \bibinfo {author} {\bibfnamefont {F.}~\bibnamefont {Colauto}},
  \bibinfo {author} {\bibfnamefont {A.~A. M.~d.}\ \bibnamefont {Oliveira}},
  \bibinfo {author} {\bibfnamefont {A.~M. H.~d.}\ \bibnamefont {Andrade}},
  \bibinfo {author} {\bibfnamefont {T.~H.}\ \bibnamefont {Johansen}}, \bibinfo
  {author} {\bibfnamefont {W.~A.}\ \bibnamefont {Ortiz}}, \ and\ \bibinfo
  {author} {\bibfnamefont {M.}~\bibnamefont {Motta}},\ }\href@noop {}
  {\bibfield  {journal} {\bibinfo  {journal} {Materials}\ }\textbf {\bibinfo
  {volume} {14}},\ \bibinfo {pages} {7274} (\bibinfo {year}
  {2021})}\BibitemShut {NoStop}%
\bibitem [{\citenamefont {Brisbois}\ \emph {et~al.}(2016)\citenamefont
  {Brisbois}, \citenamefont {Adami}, \citenamefont {Avila}, \citenamefont
  {Motta}, \citenamefont {Ortiz}, \citenamefont {Nguyen}, \citenamefont
  {Vanderbemden}, \citenamefont {Vanderheyden}, \citenamefont {Kramer},\ and\
  \citenamefont {Silhanek}}]{Brisbois2016}%
  \BibitemOpen
  \bibfield  {author} {\bibinfo {author} {\bibfnamefont {J.}~\bibnamefont
  {Brisbois}}, \bibinfo {author} {\bibfnamefont {O.-A.}\ \bibnamefont {Adami}},
  \bibinfo {author} {\bibfnamefont {J.~I.}\ \bibnamefont {Avila}}, \bibinfo
  {author} {\bibfnamefont {M.}~\bibnamefont {Motta}}, \bibinfo {author}
  {\bibfnamefont {W.~A.}\ \bibnamefont {Ortiz}}, \bibinfo {author}
  {\bibfnamefont {N.~D.}\ \bibnamefont {Nguyen}}, \bibinfo {author}
  {\bibfnamefont {P.}~\bibnamefont {Vanderbemden}}, \bibinfo {author}
  {\bibfnamefont {B.}~\bibnamefont {Vanderheyden}}, \bibinfo {author}
  {\bibfnamefont {R.~B.~G.}\ \bibnamefont {Kramer}}, \ and\ \bibinfo {author}
  {\bibfnamefont {A.~V.}\ \bibnamefont {Silhanek}},\ }\href {\doibase
  10.1103/PhysRevB.93.054521} {\bibfield  {journal} {\bibinfo  {journal} {Phys.
  Rev. B}\ }\textbf {\bibinfo {volume} {93}},\ \bibinfo {pages} {054521}
  (\bibinfo {year} {2016})}\BibitemShut {NoStop}%
\bibitem [{\citenamefont {Pinheiro}\ \emph {et~al.}(2019)\citenamefont
  {Pinheiro}, \citenamefont {Motta}, \citenamefont {Colauto}, \citenamefont
  {Johansen}, \citenamefont {Bellingeri}, \citenamefont {Bernini},
  \citenamefont {Ferdeghini},\ and\ \citenamefont {Ortiz}}]{pinheiro2019}%
  \BibitemOpen
  \bibfield  {author} {\bibinfo {author} {\bibfnamefont {L.~B. L.~G.}\
  \bibnamefont {Pinheiro}}, \bibinfo {author} {\bibfnamefont {M.}~\bibnamefont
  {Motta}}, \bibinfo {author} {\bibfnamefont {F.}~\bibnamefont {Colauto}},
  \bibinfo {author} {\bibfnamefont {T.~H.}\ \bibnamefont {Johansen}}, \bibinfo
  {author} {\bibfnamefont {E.}~\bibnamefont {Bellingeri}}, \bibinfo {author}
  {\bibfnamefont {C.}~\bibnamefont {Bernini}}, \bibinfo {author} {\bibfnamefont
  {C.}~\bibnamefont {Ferdeghini}}, \ and\ \bibinfo {author} {\bibfnamefont
  {W.~A.}\ \bibnamefont {Ortiz}},\ }\href {\doibase 10.1109/TASC.2019.2902788}
  {\bibfield  {journal} {\bibinfo  {journal} {IEEE Trans. Appl. Supercond.}\
  }\textbf {\bibinfo {volume} {29}},\ \bibinfo {pages} {1} (\bibinfo {year}
  {2019})}\BibitemShut {NoStop}%
\bibitem [{\citenamefont {Brandt}\ \emph {et~al.}(1993)\citenamefont {Brandt},
  \citenamefont {Indenbom},\ and\ \citenamefont {Forkl}}]{brandt1993type}%
  \BibitemOpen
  \bibfield  {author} {\bibinfo {author} {\bibfnamefont {E.~H.}\ \bibnamefont
  {Brandt}}, \bibinfo {author} {\bibfnamefont {M.~V.}\ \bibnamefont
  {Indenbom}}, \ and\ \bibinfo {author} {\bibfnamefont {A.}~\bibnamefont
  {Forkl}},\ }\href@noop {} {\bibfield  {journal} {\bibinfo  {journal}
  {Europhys. Lett.}\ }\textbf {\bibinfo {volume} {22}},\ \bibinfo {pages} {735}
  (\bibinfo {year} {1993})}\BibitemShut {NoStop}%
\bibitem [{\citenamefont {Foltyn}\ \emph {et~al.}(1993)\citenamefont {Foltyn},
  \citenamefont {Tiwari}, \citenamefont {Dye}, \citenamefont {Le},\ and\
  \citenamefont {Wu}}]{Foltyn1993}%
  \BibitemOpen
  \bibfield  {author} {\bibinfo {author} {\bibfnamefont {S.~R.}\ \bibnamefont
  {Foltyn}}, \bibinfo {author} {\bibfnamefont {P.}~\bibnamefont {Tiwari}},
  \bibinfo {author} {\bibfnamefont {R.~C.}\ \bibnamefont {Dye}}, \bibinfo
  {author} {\bibfnamefont {M.~Q.}\ \bibnamefont {Le}}, \ and\ \bibinfo {author}
  {\bibfnamefont {X.~D.}\ \bibnamefont {Wu}},\ }\href {\doibase
  10.1063/1.110653} {\bibfield  {journal} {\bibinfo  {journal} {Appl. Phys.
  Lett.}\ }\textbf {\bibinfo {volume} {63}},\ \bibinfo {pages} {1848} (\bibinfo
  {year} {1993})}\BibitemShut {NoStop}%
\bibitem [{\citenamefont {Huebener}\ and\ \citenamefont
  {Seher}(1969)}]{Huebener1969}%
  \BibitemOpen
  \bibfield  {author} {\bibinfo {author} {\bibfnamefont {R.~P.}\ \bibnamefont
  {Huebener}}\ and\ \bibinfo {author} {\bibfnamefont {A.}~\bibnamefont
  {Seher}},\ }\href {\doibase 10.1103/PhysRev.181.710} {\bibfield  {journal}
  {\bibinfo  {journal} {Phys. Rev.}\ }\textbf {\bibinfo {volume} {181}},\
  \bibinfo {pages} {710} (\bibinfo {year} {1969})}\BibitemShut {NoStop}%
\bibitem [{\citenamefont {Poole-Jr}\ \emph {et~al.}(2007)\citenamefont
  {Poole-Jr}, \citenamefont {Farach}, \citenamefont {Creswick},\ and\
  \citenamefont {Prozorov}}]{Poole}%
  \BibitemOpen
  \bibfield  {author} {\bibinfo {author} {\bibfnamefont {C.~P.}\ \bibnamefont
  {Poole-Jr}}, \bibinfo {author} {\bibfnamefont {H.~A.}\ \bibnamefont
  {Farach}}, \bibinfo {author} {\bibfnamefont {R.~J.}\ \bibnamefont
  {Creswick}}, \ and\ \bibinfo {author} {\bibfnamefont {R.}~\bibnamefont
  {Prozorov}},\ }\href@noop {} {\emph {\bibinfo {title} {Superconductivity}}},\
  \bibinfo {edition} {second edition}\ ed.,\ Vol.~\bibinfo {volume} {1}\
  (\bibinfo  {publisher} {Academic Press},\ \bibinfo {address} {The
  Netherlands},\ \bibinfo {year} {2007})\ p.\ \bibinfo {pages}
  {646}\BibitemShut {NoStop}%
\bibitem [{\citenamefont {Jooss}\ \emph {et~al.}(2002)\citenamefont {Jooss},
  \citenamefont {Albrecht}, \citenamefont {Kuhn}, \citenamefont {Leonhardt},\
  and\ \citenamefont {Kronm{\"u}ller}}]{jooss2002magneto}%
  \BibitemOpen
  \bibfield  {author} {\bibinfo {author} {\bibfnamefont {C.}~\bibnamefont
  {Jooss}}, \bibinfo {author} {\bibfnamefont {J.}~\bibnamefont {Albrecht}},
  \bibinfo {author} {\bibfnamefont {H.}~\bibnamefont {Kuhn}}, \bibinfo {author}
  {\bibfnamefont {S.}~\bibnamefont {Leonhardt}}, \ and\ \bibinfo {author}
  {\bibfnamefont {H.}~\bibnamefont {Kronm{\"u}ller}},\ }\href@noop {}
  {\bibfield  {journal} {\bibinfo  {journal} {Rep. Prog. Phys.}\ }\textbf
  {\bibinfo {volume} {65}},\ \bibinfo {pages} {651} (\bibinfo {year}
  {2002})}\BibitemShut {NoStop}%
\bibitem [{\citenamefont {Strongin}\ \emph {et~al.}(1970)\citenamefont
  {Strongin}, \citenamefont {Thompson}, \citenamefont {Kammerer},\ and\
  \citenamefont {Crow}}]{Strongin1970}%
  \BibitemOpen
  \bibfield  {author} {\bibinfo {author} {\bibfnamefont {M.}~\bibnamefont
  {Strongin}}, \bibinfo {author} {\bibfnamefont {R.~S.}\ \bibnamefont
  {Thompson}}, \bibinfo {author} {\bibfnamefont {O.~F.}\ \bibnamefont
  {Kammerer}}, \ and\ \bibinfo {author} {\bibfnamefont {J.~E.}\ \bibnamefont
  {Crow}},\ }\href {\doibase 10.1103/PhysRevB.1.1078} {\bibfield  {journal}
  {\bibinfo  {journal} {Phys. Rev. B}\ }\textbf {\bibinfo {volume} {1}},\
  \bibinfo {pages} {1078} (\bibinfo {year} {1970})}\BibitemShut {NoStop}%
\bibitem [{\citenamefont {Ivry}\ \emph {et~al.}(2014)\citenamefont {Ivry},
  \citenamefont {Kim}, \citenamefont {Dane}, \citenamefont {De~Fazio},
  \citenamefont {McCaughan}, \citenamefont {Sunter}, \citenamefont {Zhao},\
  and\ \citenamefont {Berggren}}]{Ivry2014}%
  \BibitemOpen
  \bibfield  {author} {\bibinfo {author} {\bibfnamefont {Y.}~\bibnamefont
  {Ivry}}, \bibinfo {author} {\bibfnamefont {C.-S.}\ \bibnamefont {Kim}},
  \bibinfo {author} {\bibfnamefont {A.~E.}\ \bibnamefont {Dane}}, \bibinfo
  {author} {\bibfnamefont {D.}~\bibnamefont {De~Fazio}}, \bibinfo {author}
  {\bibfnamefont {A.~N.}\ \bibnamefont {McCaughan}}, \bibinfo {author}
  {\bibfnamefont {K.~A.}\ \bibnamefont {Sunter}}, \bibinfo {author}
  {\bibfnamefont {Q.}~\bibnamefont {Zhao}}, \ and\ \bibinfo {author}
  {\bibfnamefont {K.~K.}\ \bibnamefont {Berggren}},\ }\href {\doibase
  10.1103/PhysRevB.90.214515} {\bibfield  {journal} {\bibinfo  {journal} {Phys.
  Rev. B}\ }\textbf {\bibinfo {volume} {90}},\ \bibinfo {pages} {214515}
  (\bibinfo {year} {2014})}\BibitemShut {NoStop}%
\bibitem [{\citenamefont {Harper}\ and\ \citenamefont
  {Tinkham}(1968)}]{HarperTinkham1968}%
  \BibitemOpen
  \bibfield  {author} {\bibinfo {author} {\bibfnamefont {F.~E.}\ \bibnamefont
  {Harper}}\ and\ \bibinfo {author} {\bibfnamefont {M.}~\bibnamefont
  {Tinkham}},\ }\href {\doibase 10.1103/PhysRev.172.441} {\bibfield  {journal}
  {\bibinfo  {journal} {Phys. Rev.}\ }\textbf {\bibinfo {volume} {172}},\
  \bibinfo {pages} {441} (\bibinfo {year} {1968})}\BibitemShut {NoStop}%
\bibitem [{\citenamefont {Tinkham}(1963)}]{Tinkham1963}%
  \BibitemOpen
  \bibfield  {author} {\bibinfo {author} {\bibfnamefont {M.}~\bibnamefont
  {Tinkham}},\ }\href {\doibase 10.1103/PhysRev.129.2413} {\bibfield  {journal}
  {\bibinfo  {journal} {Phys. Rev.}\ }\textbf {\bibinfo {volume} {129}},\
  \bibinfo {pages} {2413} (\bibinfo {year} {1963})}\BibitemShut {NoStop}%
\bibitem [{\citenamefont {Cody}\ and\ \citenamefont
  {Miller}(1968)}]{CodyMiller1968}%
  \BibitemOpen
  \bibfield  {author} {\bibinfo {author} {\bibfnamefont {G.~D.}\ \bibnamefont
  {Cody}}\ and\ \bibinfo {author} {\bibfnamefont {R.~E.}\ \bibnamefont
  {Miller}},\ }\href {\doibase 10.1103/PhysRev.173.481} {\bibfield  {journal}
  {\bibinfo  {journal} {Phys. Rev.}\ }\textbf {\bibinfo {volume} {173}},\
  \bibinfo {pages} {481} (\bibinfo {year} {1968})}\BibitemShut {NoStop}%
\bibitem [{\citenamefont {Takayama}\ \emph {et~al.}(1971)\citenamefont
  {Takayama}, \citenamefont {Ōgushi},\ and\ \citenamefont
  {Shibuya}}]{Takayama1971}%
  \BibitemOpen
  \bibfield  {author} {\bibinfo {author} {\bibfnamefont {T.}~\bibnamefont
  {Takayama}}, \bibinfo {author} {\bibfnamefont {T.}~\bibnamefont {Ōgushi}}, \
  and\ \bibinfo {author} {\bibfnamefont {Y.}~\bibnamefont {Shibuya}},\ }\href
  {\doibase 10.1143/JPSJ.30.1083} {\bibfield  {journal} {\bibinfo  {journal}
  {J. Phys. Soc. Jpn.}\ }\textbf {\bibinfo {volume} {30}},\ \bibinfo {pages}
  {1083} (\bibinfo {year} {1971})}\BibitemShut {NoStop}%
\bibitem [{\citenamefont {Brandt}\ \emph {et~al.}(1971)\citenamefont {Brandt},
  \citenamefont {Parks},\ and\ \citenamefont {Chaudhari}}]{Brandt1971}%
  \BibitemOpen
  \bibfield  {author} {\bibinfo {author} {\bibfnamefont {B.~L.}\ \bibnamefont
  {Brandt}}, \bibinfo {author} {\bibfnamefont {R.~D.}\ \bibnamefont {Parks}}, \
  and\ \bibinfo {author} {\bibfnamefont {R.~D.}\ \bibnamefont {Chaudhari}},\
  }\href {\doibase 10.1007/BF00628436} {\bibfield  {journal} {\bibinfo
  {journal} {J. Low Temp. Phys.}\ }\textbf {\bibinfo {volume} {4}},\ \bibinfo
  {pages} {41} (\bibinfo {year} {1971})}\BibitemShut {NoStop}%
\bibitem [{\citenamefont {Albrecht}\ \emph {et~al.}(2007)\citenamefont
  {Albrecht}, \citenamefont {Matveev}, \citenamefont {Strempfer}, \citenamefont
  {Habermeier}, \citenamefont {Shantsev}, \citenamefont {Galperin},\ and\
  \citenamefont {Johansen}}]{albrechtprl2007}%
  \BibitemOpen
  \bibfield  {author} {\bibinfo {author} {\bibfnamefont {J.}~\bibnamefont
  {Albrecht}}, \bibinfo {author} {\bibfnamefont {A.~T.}\ \bibnamefont
  {Matveev}}, \bibinfo {author} {\bibfnamefont {J.}~\bibnamefont {Strempfer}},
  \bibinfo {author} {\bibfnamefont {H.-U.}\ \bibnamefont {Habermeier}},
  \bibinfo {author} {\bibfnamefont {D.~V.}\ \bibnamefont {Shantsev}}, \bibinfo
  {author} {\bibfnamefont {Y.~M.}\ \bibnamefont {Galperin}}, \ and\ \bibinfo
  {author} {\bibfnamefont {T.~H.}\ \bibnamefont {Johansen}},\ }\href {\doibase
  10.1103/PhysRevLett.98.117001} {\bibfield  {journal} {\bibinfo  {journal}
  {Phys. Rev. Lett.}\ }\textbf {\bibinfo {volume} {98}},\ \bibinfo {pages}
  {117001} (\bibinfo {year} {2007})}\BibitemShut {NoStop}%
\bibitem [{\citenamefont {Jing}\ \emph {et~al.}(2016)\citenamefont {Jing},
  \citenamefont {Yong},\ and\ \citenamefont {Zhou}}]{JingZhou2016}%
  \BibitemOpen
  \bibfield  {author} {\bibinfo {author} {\bibfnamefont {Z.}~\bibnamefont
  {Jing}}, \bibinfo {author} {\bibfnamefont {H.}~\bibnamefont {Yong}}, \ and\
  \bibinfo {author} {\bibfnamefont {Y.}~\bibnamefont {Zhou}},\ }\href {\doibase
  10.1088/0953-2048/29/10/105001} {\bibfield  {journal} {\bibinfo  {journal}
  {Supercond. Sci. Technol.}\ }\textbf {\bibinfo {volume} {29}},\ \bibinfo
  {pages} {105001} (\bibinfo {year} {2016})}\BibitemShut {NoStop}%
\bibitem [{\citenamefont {Johansen}\ \emph {et~al.}(2002)\citenamefont
  {Johansen}, \citenamefont {Baziljevich}, \citenamefont {Shantsev},
  \citenamefont {Goa}, \citenamefont {pe~rin}, \citenamefont {Kang},
  \citenamefont {Kim}, \citenamefont {Choi}, \citenamefont {Kim},\ and\
  \citenamefont {Lee}}]{Johansen2002}%
  \BibitemOpen
  \bibfield  {author} {\bibinfo {author} {\bibfnamefont {T.~H.}\ \bibnamefont
  {Johansen}}, \bibinfo {author} {\bibfnamefont {M.}~\bibnamefont
  {Baziljevich}}, \bibinfo {author} {\bibfnamefont {D.~V.}\ \bibnamefont
  {Shantsev}}, \bibinfo {author} {\bibfnamefont {P.~E.}\ \bibnamefont {Goa}},
  \bibinfo {author} {\bibfnamefont {Y.~M.~G.}\ \bibnamefont {pe~rin}}, \bibinfo
  {author} {\bibfnamefont {W.~N.}\ \bibnamefont {Kang}}, \bibinfo {author}
  {\bibfnamefont {H.~J.}\ \bibnamefont {Kim}}, \bibinfo {author} {\bibfnamefont
  {E.~M.}\ \bibnamefont {Choi}}, \bibinfo {author} {\bibfnamefont {M.-S.}\
  \bibnamefont {Kim}}, \ and\ \bibinfo {author} {\bibfnamefont {S.~I.}\
  \bibnamefont {Lee}},\ }\href {\doibase 10.1209/epl/i2002-00146-1} {\bibfield
  {journal} {\bibinfo  {journal} {Europhys. Lett.}\ }\textbf {\bibinfo {volume}
  {59}},\ \bibinfo {pages} {599} (\bibinfo {year} {2002})}\BibitemShut
  {NoStop}%
\bibitem [{\citenamefont {Welling}\ \emph {et~al.}(2004)\citenamefont
  {Welling}, \citenamefont {Westerwaal}, \citenamefont {Lohstroh},\ and\
  \citenamefont {Wijngaarden}}]{Welling2004}%
  \BibitemOpen
  \bibfield  {author} {\bibinfo {author} {\bibfnamefont {M.~S.}\ \bibnamefont
  {Welling}}, \bibinfo {author} {\bibfnamefont {R.~J.}\ \bibnamefont
  {Westerwaal}}, \bibinfo {author} {\bibfnamefont {W.}~\bibnamefont
  {Lohstroh}}, \ and\ \bibinfo {author} {\bibfnamefont {R.~J.}\ \bibnamefont
  {Wijngaarden}},\ }\href {\doibase
  https://doi.org/10.1016/j.physc.2004.06.011} {\bibfield  {journal} {\bibinfo
  {journal} {Physica C}\ }\textbf {\bibinfo {volume} {411}},\ \bibinfo {pages}
  {11} (\bibinfo {year} {2004})}\BibitemShut {NoStop}%
\bibitem [{\citenamefont {Vestg{\aa}rden}\ \emph {et~al.}(2013)\citenamefont
  {Vestg{\aa}rden}, \citenamefont {Shantsev}, \citenamefont {Galperin},\ and\
  \citenamefont {Johansen}}]{Vestgarden2013SUST}%
  \BibitemOpen
  \bibfield  {author} {\bibinfo {author} {\bibfnamefont {J.~I.}\ \bibnamefont
  {Vestg{\aa}rden}}, \bibinfo {author} {\bibfnamefont {D.~V.}\ \bibnamefont
  {Shantsev}}, \bibinfo {author} {\bibfnamefont {Y.~M.}\ \bibnamefont
  {Galperin}}, \ and\ \bibinfo {author} {\bibfnamefont {T.~H.}\ \bibnamefont
  {Johansen}},\ }\href {\doibase 10.1088/0953-2048/26/5/055012} {\enquote
  {\bibinfo {title} {The diversity of flux avalanche patterns in
  superconducting films},}\ } (\bibinfo {year} {2013})\BibitemShut {NoStop}%
\bibitem [{\citenamefont {Johansen}\ \emph {et~al.}(2001)\citenamefont
  {Johansen}, \citenamefont {Baziljevich}, \citenamefont {Shantsev},
  \citenamefont {Goa}, \citenamefont {Galperin}, \citenamefont {Kang},
  \citenamefont {Kim}, \citenamefont {Choi}, \citenamefont {Kim},\ and\
  \citenamefont {Lee}}]{Johansen2001}%
  \BibitemOpen
  \bibfield  {author} {\bibinfo {author} {\bibfnamefont {T.~H.}\ \bibnamefont
  {Johansen}}, \bibinfo {author} {\bibfnamefont {M.}~\bibnamefont
  {Baziljevich}}, \bibinfo {author} {\bibfnamefont {D.~V.}\ \bibnamefont
  {Shantsev}}, \bibinfo {author} {\bibfnamefont {P.~E.}\ \bibnamefont {Goa}},
  \bibinfo {author} {\bibfnamefont {Y.~M.}\ \bibnamefont {Galperin}}, \bibinfo
  {author} {\bibfnamefont {W.~N.}\ \bibnamefont {Kang}}, \bibinfo {author}
  {\bibfnamefont {H.~J.}\ \bibnamefont {Kim}}, \bibinfo {author} {\bibfnamefont
  {E.~M.}\ \bibnamefont {Choi}}, \bibinfo {author} {\bibfnamefont {M.~S.}\
  \bibnamefont {Kim}}, \ and\ \bibinfo {author} {\bibfnamefont {S.~I.}\
  \bibnamefont {Lee}},\ }\href {\doibase 10.1088/0953-2048/14/9/319} {\bibfield
   {journal} {\bibinfo  {journal} {Supercond. Sci. Technol.}\ }\textbf
  {\bibinfo {volume} {14}},\ \bibinfo {pages} {726} (\bibinfo {year}
  {2001})}\BibitemShut {NoStop}%
\bibitem [{\citenamefont {Choi}\ \emph {et~al.}(2008)\citenamefont {Choi},
  \citenamefont {Yurchenko}, \citenamefont {Johansen}, \citenamefont {Lee},
  \citenamefont {Lee}, \citenamefont {Kang},\ and\ \citenamefont
  {Lee}}]{Choi2008}%
  \BibitemOpen
  \bibfield  {author} {\bibinfo {author} {\bibfnamefont {E.-M.}\ \bibnamefont
  {Choi}}, \bibinfo {author} {\bibfnamefont {V.~V.}\ \bibnamefont {Yurchenko}},
  \bibinfo {author} {\bibfnamefont {T.~H.}\ \bibnamefont {Johansen}}, \bibinfo
  {author} {\bibfnamefont {H.-S.}\ \bibnamefont {Lee}}, \bibinfo {author}
  {\bibfnamefont {J.~Y.}\ \bibnamefont {Lee}}, \bibinfo {author} {\bibfnamefont
  {W.~N.}\ \bibnamefont {Kang}}, \ and\ \bibinfo {author} {\bibfnamefont
  {S.-I.}\ \bibnamefont {Lee}},\ }\href {\doibase
  10.1088/0953-2048/22/1/015011} {\bibfield  {journal} {\bibinfo  {journal}
  {Supercond. Sci. Technol.}\ }\textbf {\bibinfo {volume} {22}},\ \bibinfo
  {pages} {015011} (\bibinfo {year} {2008})}\BibitemShut {NoStop}%
\bibitem [{\citenamefont {Pinheiro}\ \emph {et~al.}(2020)\citenamefont
  {Pinheiro}, \citenamefont {Caputo}, \citenamefont {Cirillo}, \citenamefont
  {Attanasio}, \citenamefont {Johansen}, \citenamefont {Ortiz}, \citenamefont
  {Silhanek},\ and\ \citenamefont {Motta}}]{Pinheiro2020}%
  \BibitemOpen
  \bibfield  {author} {\bibinfo {author} {\bibfnamefont {L.~B.}\ \bibnamefont
  {Pinheiro}}, \bibinfo {author} {\bibfnamefont {M.}~\bibnamefont {Caputo}},
  \bibinfo {author} {\bibfnamefont {C.}~\bibnamefont {Cirillo}}, \bibinfo
  {author} {\bibfnamefont {C.}~\bibnamefont {Attanasio}}, \bibinfo {author}
  {\bibfnamefont {T.~H.}\ \bibnamefont {Johansen}}, \bibinfo {author}
  {\bibfnamefont {W.~A.}\ \bibnamefont {Ortiz}}, \bibinfo {author}
  {\bibfnamefont {A.~V.}\ \bibnamefont {Silhanek}}, \ and\ \bibinfo {author}
  {\bibfnamefont {M.}~\bibnamefont {Motta}},\ }\href {\doibase
  10.1063/10.0000868} {\bibfield  {journal} {\bibinfo  {journal} {Low Temp.
  Phys.}\ }\textbf {\bibinfo {volume} {46}},\ \bibinfo {pages} {365} (\bibinfo
  {year} {2020})}\BibitemShut {NoStop}%
\bibitem [{\citenamefont {Qviller}\ \emph {et~al.}(2010)\citenamefont
  {Qviller}, \citenamefont {Yurchenko}, \citenamefont {Eliassen}, \citenamefont
  {rden}, \citenamefont {Johansen}, \citenamefont {Nevala}, \citenamefont
  {Maasilta}, \citenamefont {Senapati},\ and\ \citenamefont
  {Budhani}}]{Qviller2010}%
  \BibitemOpen
  \bibfield  {author} {\bibinfo {author} {\bibfnamefont {A.~J.}\ \bibnamefont
  {Qviller}}, \bibinfo {author} {\bibfnamefont {V.~V.}\ \bibnamefont
  {Yurchenko}}, \bibinfo {author} {\bibfnamefont {K.}~\bibnamefont {Eliassen}},
  \bibinfo {author} {\bibfnamefont {J.~I.~V.}\ \bibnamefont {rden}}, \bibinfo
  {author} {\bibfnamefont {T.~H.}\ \bibnamefont {Johansen}}, \bibinfo {author}
  {\bibfnamefont {M.~R.}\ \bibnamefont {Nevala}}, \bibinfo {author}
  {\bibfnamefont {I.~J.}\ \bibnamefont {Maasilta}}, \bibinfo {author}
  {\bibfnamefont {K.}~\bibnamefont {Senapati}}, \ and\ \bibinfo {author}
  {\bibfnamefont {R.~C.}\ \bibnamefont {Budhani}},\ }\href {\doibase
  https://doi.org/10.1016/j.physc.2010.02.066} {\bibfield  {journal} {\bibinfo
  {journal} {Physica C}\ }\textbf {\bibinfo {volume} {470}},\ \bibinfo {pages}
  {897} (\bibinfo {year} {2010})}\BibitemShut {NoStop}%
\bibitem [{\citenamefont {Smith}(1995)}]{smith1995thin}%
  \BibitemOpen
  \bibfield  {author} {\bibinfo {author} {\bibfnamefont {D.~L.}\ \bibnamefont
  {Smith}},\ }\href {https://books.google.com.br/books?id=kTVkwRWwxfYC} {\emph
  {\bibinfo {title} {Thin-Film Deposition: Principles and Practice}}}\
  (\bibinfo  {publisher} {McGraw-Hill Education},\ \bibinfo {year}
  {1995})\BibitemShut {NoStop}%
\bibitem [{\citenamefont {{Novaetech S.r.l.}}(2021)}]{openQCM}%
  \BibitemOpen
  \bibfield  {author} {\bibinfo {author} {\bibnamefont {{Novaetech S.r.l.}}},\
  }\href@noop {} {\enquote {\bibinfo {title} {Quartz crystal microbalance},}\
  }\bibinfo {howpublished} {\url{https://openqcm.com/openqcm}} (\bibinfo {year}
  {2021}),\ \bibinfo {note} {[Online; accessed 8-December-2021]}\BibitemShut
  {NoStop}%
\bibitem [{\citenamefont {{Walker, P. and Tarn, W.
  H.}}(1990)}]{metaletchants1990}%
  \BibitemOpen
  \bibfield  {author} {\bibinfo {author} {\bibnamefont {{Walker, P. and Tarn,
  W. H.}}},\ }\href {\doibase 10.1201/9780367803087} {\emph {\bibinfo {title}
  {{CRC Handbook of Metal Etchants}}}}\ (\bibinfo  {publisher} {CRC {P}ress},\
  \bibinfo {year} {1990})\ p.\ \bibinfo {pages} {1415}\BibitemShut {NoStop}%
\bibitem [{\citenamefont {Perrone}\ \emph {et~al.}(2013)\citenamefont
  {Perrone}, \citenamefont {Gontad}, \citenamefont {Lorusso}, \citenamefont
  {{Di Giulio}}, \citenamefont {Broitman},\ and\ \citenamefont
  {Ferrario}}]{Perrone2013Pb}%
  \BibitemOpen
  \bibfield  {author} {\bibinfo {author} {\bibfnamefont {A.}~\bibnamefont
  {Perrone}}, \bibinfo {author} {\bibfnamefont {F.}~\bibnamefont {Gontad}},
  \bibinfo {author} {\bibfnamefont {A.}~\bibnamefont {Lorusso}}, \bibinfo
  {author} {\bibfnamefont {M.}~\bibnamefont {{Di Giulio}}}, \bibinfo {author}
  {\bibfnamefont {E.}~\bibnamefont {Broitman}}, \ and\ \bibinfo {author}
  {\bibfnamefont {M.}~\bibnamefont {Ferrario}},\ }\href {\doibase
  https://doi.org/10.1016/j.nima.2013.07.082} {\bibfield  {journal} {\bibinfo
  {journal} {Nucl. Instrum. Methods Phys. Res., Sect. A}\ }\textbf {\bibinfo
  {volume} {729}},\ \bibinfo {pages} {451} (\bibinfo {year}
  {2013})}\BibitemShut {NoStop}%
\bibitem [{\citenamefont {Lorusso}\ \emph {et~al.}(2015)\citenamefont
  {Lorusso}, \citenamefont {Gontad}, \citenamefont {Broitman}, \citenamefont
  {Chiadroni},\ and\ \citenamefont {Perrone}}]{Lorusso2015Pb}%
  \BibitemOpen
  \bibfield  {author} {\bibinfo {author} {\bibfnamefont {A.}~\bibnamefont
  {Lorusso}}, \bibinfo {author} {\bibfnamefont {F.}~\bibnamefont {Gontad}},
  \bibinfo {author} {\bibfnamefont {E.}~\bibnamefont {Broitman}}, \bibinfo
  {author} {\bibfnamefont {E.}~\bibnamefont {Chiadroni}}, \ and\ \bibinfo
  {author} {\bibfnamefont {A.}~\bibnamefont {Perrone}},\ }\href {\doibase
  https://doi.org/10.1016/j.tsf.2015.02.033} {\bibfield  {journal} {\bibinfo
  {journal} {Thin Solid Films}\ }\textbf {\bibinfo {volume} {579}},\ \bibinfo
  {pages} {50} (\bibinfo {year} {2015})}\BibitemShut {NoStop}%
\bibitem [{\citenamefont {Broitman}\ \emph {et~al.}(2016)\citenamefont
  {Broitman}, \citenamefont {Flores-Ruiz}, \citenamefont {Di~Giulio},
  \citenamefont {Gontad}, \citenamefont {Lorusso},\ and\ \citenamefont
  {Perrone}}]{Broitman2016Pb}%
  \BibitemOpen
  \bibfield  {author} {\bibinfo {author} {\bibfnamefont {E.}~\bibnamefont
  {Broitman}}, \bibinfo {author} {\bibfnamefont {F.~J.}\ \bibnamefont
  {Flores-Ruiz}}, \bibinfo {author} {\bibfnamefont {M.}~\bibnamefont
  {Di~Giulio}}, \bibinfo {author} {\bibfnamefont {F.}~\bibnamefont {Gontad}},
  \bibinfo {author} {\bibfnamefont {A.}~\bibnamefont {Lorusso}}, \ and\
  \bibinfo {author} {\bibfnamefont {A.}~\bibnamefont {Perrone}},\ }\href
  {\doibase 10.1116/1.4936080} {\bibfield  {journal} {\bibinfo  {journal} {J.
  Vac. Sci. Technol. A}\ }\textbf {\bibinfo {volume} {34}},\ \bibinfo {pages}
  {021505} (\bibinfo {year} {2016})}\BibitemShut {NoStop}%
\bibitem [{\citenamefont {{Ludwig Reimer}}(1985)}]{Reimer1985}%
  \BibitemOpen
  \bibfield  {author} {\bibinfo {author} {\bibnamefont {{Ludwig Reimer}}},\
  }\href {https://doi.org/10.1007/978-3-662-13562-4_3} {\emph {\bibinfo {title}
  {{ Electron Scattering and Diffusion}}}}\ (\bibinfo  {publisher} {Springer},\
  \bibinfo {address} {Berlin, Heidelberg},\ \bibinfo {year} {1985})\ pp.\
  \bibinfo {pages} {1--388}\BibitemShut {NoStop}%
\bibitem [{\citenamefont {Pascual}\ \emph {et~al.}(1990)\citenamefont
  {Pascual}, \citenamefont {Cruz}, \citenamefont {Ferreira},\ and\
  \citenamefont {Gomes}}]{Pascual1990edsthick}%
  \BibitemOpen
  \bibfield  {author} {\bibinfo {author} {\bibfnamefont {R.}~\bibnamefont
  {Pascual}}, \bibinfo {author} {\bibfnamefont {L.~R.}\ \bibnamefont {Cruz}},
  \bibinfo {author} {\bibfnamefont {C.~L.}\ \bibnamefont {Ferreira}}, \ and\
  \bibinfo {author} {\bibfnamefont {D.~T.}\ \bibnamefont {Gomes}},\ }\href
  {\doibase https://doi.org/10.1016/0040-6090(90)90092-R} {\bibfield  {journal}
  {\bibinfo  {journal} {Thin Solid Films}\ }\textbf {\bibinfo {volume} {185}},\
  \bibinfo {pages} {279} (\bibinfo {year} {1990})}\BibitemShut {NoStop}%
\bibitem [{\citenamefont {Ng}\ \emph {et~al.}(2006)\citenamefont {Ng},
  \citenamefont {Wei}, \citenamefont {Lai},\ and\ \citenamefont
  {Goh}}]{Ng2006edsthick}%
  \BibitemOpen
  \bibfield  {author} {\bibinfo {author} {\bibfnamefont {F.~L.}\ \bibnamefont
  {Ng}}, \bibinfo {author} {\bibfnamefont {J.}~\bibnamefont {Wei}}, \bibinfo
  {author} {\bibfnamefont {F.~K.}\ \bibnamefont {Lai}}, \ and\ \bibinfo
  {author} {\bibfnamefont {K.~L.}\ \bibnamefont {Goh}},\ }\href@noop {}
  {\bibfield  {journal} {\bibinfo  {journal} {Appl. Surf. Sci.}\ }\textbf
  {\bibinfo {volume} {252}},\ \bibinfo {pages} {3972} (\bibinfo {year}
  {2006})}\BibitemShut {NoStop}%
\bibitem [{\citenamefont {Zhuang}\ \emph {et~al.}(2009)\citenamefont {Zhuang},
  \citenamefont {Bao}, \citenamefont {Wang}, \citenamefont {Li}, \citenamefont
  {Ma},\ and\ \citenamefont {Lv}}]{Zhuang2009edsthick}%
  \BibitemOpen
  \bibfield  {author} {\bibinfo {author} {\bibfnamefont {L.}~\bibnamefont
  {Zhuang}}, \bibinfo {author} {\bibfnamefont {S.}~\bibnamefont {Bao}},
  \bibinfo {author} {\bibfnamefont {R.}~\bibnamefont {Wang}}, \bibinfo {author}
  {\bibfnamefont {S.}~\bibnamefont {Li}}, \bibinfo {author} {\bibfnamefont
  {L.}~\bibnamefont {Ma}}, \ and\ \bibinfo {author} {\bibfnamefont
  {D.}~\bibnamefont {Lv}},\ }in\ \href@noop {} {\emph {\bibinfo {booktitle}
  {2009 International Conference on Applied Superconductivity and
  Electromagnetic Devices}}}\ (\bibinfo {organization} {IEEE},\ \bibinfo {year}
  {2009})\ pp.\ \bibinfo {pages} {142--144}\BibitemShut {NoStop}%
\bibitem [{\citenamefont {Habiger}\ and\ \citenamefont
  {Stein}(1992)}]{Habiger1992}%
  \BibitemOpen
  \bibfield  {author} {\bibinfo {author} {\bibfnamefont {K.~W.}\ \bibnamefont
  {Habiger}}\ and\ \bibinfo {author} {\bibfnamefont {C.}~\bibnamefont
  {Stein}},\ }\href {\doibase https://doi.org/10.1016/0040-6090(92)90710-S}
  {\bibfield  {journal} {\bibinfo  {journal} {Thin Solid Films}\ }\textbf
  {\bibinfo {volume} {215}},\ \bibinfo {pages} {108} (\bibinfo {year}
  {1992})}\BibitemShut {NoStop}%
\bibitem [{\citenamefont {Goldstein}\ \emph {et~al.}(2003)\citenamefont
  {Goldstein}, \citenamefont {Newbury}, \citenamefont {Echlin}, \citenamefont
  {Joy}, \citenamefont {Lyman}, \citenamefont {Lifshin}, \citenamefont
  {Sawyer},\ and\ \citenamefont {Michael}}]{Goldstein2003_analy}%
  \BibitemOpen
  \bibfield  {author} {\bibinfo {author} {\bibfnamefont {J.~I.}\ \bibnamefont
  {Goldstein}}, \bibinfo {author} {\bibfnamefont {D.~E.}\ \bibnamefont
  {Newbury}}, \bibinfo {author} {\bibfnamefont {P.}~\bibnamefont {Echlin}},
  \bibinfo {author} {\bibfnamefont {D.~C.}\ \bibnamefont {Joy}}, \bibinfo
  {author} {\bibfnamefont {C.~E.}\ \bibnamefont {Lyman}}, \bibinfo {author}
  {\bibfnamefont {E.}~\bibnamefont {Lifshin}}, \bibinfo {author} {\bibfnamefont
  {L.}~\bibnamefont {Sawyer}}, \ and\ \bibinfo {author} {\bibfnamefont {J.~R.}\
  \bibnamefont {Michael}},\ }\href {\doibase 10.1007/978-1-4615-0215-9} {\emph
  {\bibinfo {title} {Scanning Electron Microscopy and X-Ray Microanalysis}}},\
  \bibinfo {edition} {3rd}\ ed.\ (\bibinfo  {publisher} {Springer Science},\
  \bibinfo {address} {New York},\ \bibinfo {year} {2003})\BibitemShut {NoStop}%
\bibitem [{\citenamefont {Reed}(2005)}]{Reed2005m_analy}%
  \BibitemOpen
  \bibfield  {author} {\bibinfo {author} {\bibfnamefont {S.~J.~B.}\
  \bibnamefont {Reed}},\ }\href {\doibase 10.1017/CBO9780511610561} {\emph
  {\bibinfo {title} {Electron Microprobe Analysis and Scanning Electron
  Microscopy in Geology}}},\ \bibinfo {edition} {2nd}\ ed.\ (\bibinfo
  {publisher} {Cambridge University Press},\ \bibinfo {address} {Cambridge},\
  \bibinfo {year} {2005})\BibitemShut {NoStop}%
\end{thebibliography}%

\end{document}